\documentclass[]{JHEP3}
\usepackage{graphics,amsmath,epsfig,latexsym}
\usepackage{epsfig}
\usepackage{graphicx}
\newcommand{\be}{\begin{equation}}
\newcommand{\ee}{\end{equation}}
\newcommand{\beq}{\begin{equation}}
\newcommand{\eeq}{\end{equation}}
\newcommand{\bea}{\begin{eqnarray}}
\newcommand{\eea}{\end{eqnarray}}
\newcommand{\ie}{{\it i.e.}}
\newcommand{\eg}{{\it e.g.}}
\newcommand{\np}{{}^{{\scriptscriptstyle (n+1)}}}
\newcommand{\n}{{}^{{\scriptscriptstyle (n)}}}

\newcommand{\inp}{{}^{{\scriptscriptstyle (p+1)}}}
\newcommand{\p}{{}^{{\scriptscriptstyle (p)}}}
\newcommand{\nmp}{{}^{{\scriptscriptstyle (n-p)}}}
\newcommand{\tr}{{\rm Tr}\,}
\title{Mirage resolution of cosmological singularities}
\author{Fr\'ed\'eric Leblond \\ Enrico Fermi
Institute, University of Chicago, \\ 5640 S. Ellis Av., Chicago, IL
60637, USA \\ \email{fleblond@theory.uchicago.edu}}

\abstract{We study time--dependent backgrounds in the low~energy
regimes of string theories. In particular the emphasis is on the
general study of exotic phenomena such as positive acceleration
and gravitational bounces. We generalize the usual
Hawking--Penrose cosmological singularity theorems to
higher--dimensional spacetimes and discuss their implications for
time--dependent solutions in supergravity theories. The explicit
examples we consider fall in two categories. First we consider
effective lower--dimensional gravitational theories obtained from
compactifications of ten and eleven--dimensional supergravity. We
argue and explain why non--singular solutions (\eg, with positive
acceleration and possibly a bounce) can in principle be obtained.
However we show that their uplift to higher dimensions is always
singular as predicted by the theorems. Secondly we revisit the
issue of supergravity s--branes. Our main result is to propose a
generic mechanism by which the usual singularities can be
resolved.}

\preprint{hep-th/xxxxxxx \\ EFI-04-07}

\begin{document}
\small{

\section{Introduction}

Perhaps the most pressing problem in theoretical physics is to
explain the current state of acceleration in the universe
\cite{supernovae}. Related to this is the fact that there is no
candidate theory of quantum gravity providing a consistent
mechanism associated with the generation of a positive
cosmological constant (see, however, ref.~\cite{kklt}). Another
outstanding problem consists in our inability to convincingly
resolve cosmological singularities.\footnote{However interesting
attempts were made in the context of string theory (see
refs.~\cite{moore,chicago}).} The purpose of this paper is to
explore further the nature of cosmological singularities and
positive acceleration in the low~energy limits of string theories.
This means our work is relevant in the context of the
ten--dimensional and eleven--dimensional supergravity theories.
For ten--dimensional supergravity as long as the scales involved
are considerably larger than the string length (and that the
string coupling is kept small), the corresponding low--energy
actions will capture relevant time--dependent physics. More
specifically we are considering ten-- or eleven--dimensional
Einstein gravity coupled to sources found in supergravity
theories. The latter will consist in massless fields such as the
dilaton and the Ramond--Ramond forms as well as extended sources
such as $p$--branes.

A related motivation for our work was the study of gravitational
throats. These are spacetime regions associated with a direction
along which a spatial volume element goes through a
minimum.\footnote{This simple definition lifts the restriction
that a throat must be associated with a compact sub--manifold. For
instance when considering time--dependent homogeneous and
isotropic FLRW cosmology the term gravitational throat is not only
relevant for a spherical foliation.} This is illustrated on
figure~\ref{thr} for a spherical hypersurface. If the minimum is
reached via a timelike direction the resulting geometry is a
bouncing cosmology. When the extremum is the result of a
contraction followed by an expansion along a spacelike direction
this a wormhole. \EPSFIGURE[r]{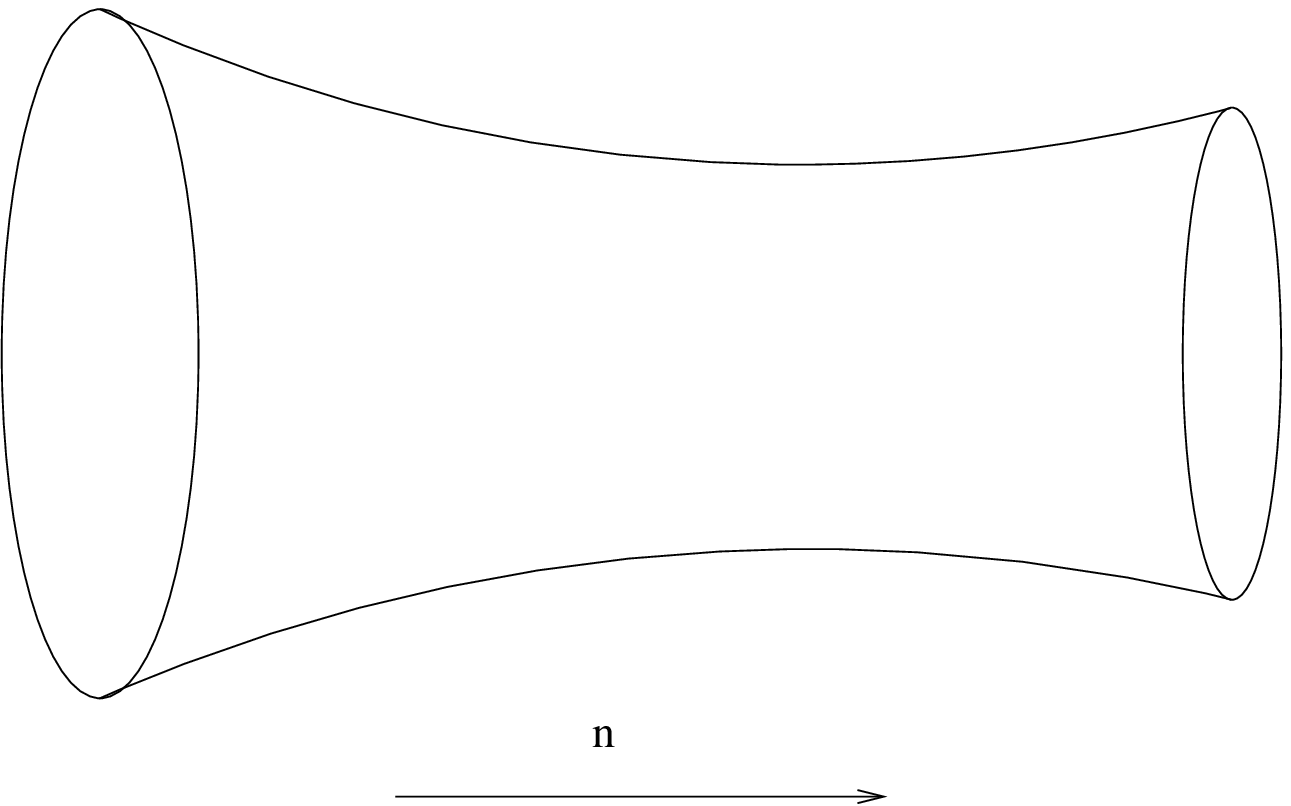,width=50mm}{Schematic
depiction of a spherical gravitational throat. If $n$ is a
timelike vector the corresponding geometry is a bouncing
cosmology. Cases for which $n$ is spacelike are associated with
Lorentzian or Euclidean wormholes.\label{thr}} Early work on
Euclidean wormholes focused on their potential role in a theory of
quantum gravity based on the Euclidean path integral formalism
(see, \eg, refs.~\cite{coleman,klebanov,giddings}). A recent
attempt \cite{maoz} (see also ref.~\cite{mcinnes}) was made to
consider the role played by these configurations in gauge/gravity
dualities. Four--dimensional Lorentzian wormhole solutions were
found in the past (see, \eg, refs.~\cite{thorne,visser}). It was
then shown that the existence of such a gravitational throat
requires that the matter supporting it violates the weak energy
condition. In section~\ref{energy} we derive and explain the
gravitational energy conditions. There are also topological
censorship theorems \cite{galloway} showing that wormholes cannot
exist unless their disconnected boundaries are separated by an
horizon.

In this paper we primarily study bouncing cosmologies. The
generalization of our results to Lorentzian wormholes is presented
in the discussion section. It is a well--known fact that
time--dependent gravitational bounces can only be supported by
matter sources violating the strong energy condition. In
section~\ref{singular} we generalize to higher dimensions the
Hawking--Penrose singularity theorems. This ensures that theories
with sources respecting the latter condition do not admit
non--singular time--dependent solutions. However we point out that
this conclusion does not necessarily hold in higher--dimensional
spacetimes where only a sub--manifold is bouncing. An example of
this would be if the geometry on this sub--space is that of
de~Sitter space in global coordinates. In section~\ref{bounces} we
consider this phenomenon in details. The gravitational theory on
the bouncing sub--manifold will appear regular but this is only an
illusion since the theorems predict that singular points must
develop as seen from the higher--dimensional geometry point of view.
Another possibility is that rather than being associated with a bounce the
sub--manifold is a non--singular forever expanding (or forever
contracting) spacetime. This is a feature associated with the representation
of de~Sitter space in inflationary coordinates. The region where the volume of the spacetime
vanishes is then a non--singular horizon.

In section~\ref{sugra} we consider homogeneous and isotropic
cosmological solutions of ($p+1$)--dimensional Einstein gravity
($p\leq 10$) coupled to a scalar field with positive exponential
potential. We find analytic solutions for spacetimes with flat
foliations. We obtain their asymptotic behavior and show these
geometries are always associated with an intermediary phase of
positive acceleration. Although the scalar field is allowed to
violate the strong energy condition we find there is always a
curvature singularity either in the past or the future when the
scale factor becomes small. Then we consider the corresponding
spacetimes with positive spatial curvature. We find interesting
non--singular bouncing cosmologies but show that if the slope
parameter for the potential is too large the geometries are
singular. Then in section~\ref{applications} we consider flux
compactifications of ten-- and eleven--dimensional supergravity on
maximally symmetric spaces. This provides a natural way to embed
the $k=0$ and $k=+1$ lower--dimensional spacetimes studied in
section~\ref{sugra}. We discuss the resulting geometries in the
context of singularity resolution from the point of view of a
lower--dimensional observer.

Finally in section~\ref{spacelikeb} we apply our results to the
study of supergravity s--branes. We use the singularity theorems
to show that s$p$--brane with $p\leq 7$ are always singular. The
s8--brane evades the theorems but we show that it is nevertheless
singular. When studying unstable branes from a gravitational perspective
we have always assumed that the end point (and the time
when brane formation starts) of the decay is the closed string
vacuum with vanishing energy density. We show that considering a
more general setup where unstable branes evolve inside larger
stable branes might generically lead to a resolution of all
previously found singularities.

\section{Background notions}
\label{background}

\subsection{Equations of motion}
\label{eom}

We begin by deriving the general relativistic equations of motion that
will be used throughout the paper. We choose to write these down
in terms of the extrinsic and intrinsic curvatures. Not only does
this simplify the analysis but it allows us to address general
issues (\eg, the role of inhomogeneity and anisotropy) which are
hard to take into account when a more specific metric ansatz is
used. Beginning with section~\ref{hubble} our analysis will be in
terms of a more intuitively accessible homogeneous metric ansatz.

The most general form for the metric of a $(n+1)$--dimensional
gravitational background is \beq \label{general} ds^2 = -(N^2 -
N_i N^i) dt^2 + 2 N_j dx^j dt + \n g_{ij} dx^i dx^j, \eeq where
$i,j=1,\, ... \, , n$ and $N=N(t,x^i)$, $N^{i}=N^{i}(t,x^i)$ are
respectively the lapse function and the shift functions. The
spacetime is assumed to be globally hyperbolic\footnote{This
assumption will be momentarily relaxed when we consider
singularity theorems in section~\ref{singular}.} with the geometry
of the constant $t$ hypersurfaces characterized by the spatial
metric $\n g_{ij}$. All information about the intrinsic curvature
of the Cauchy surfaces is contained in the Riemann tensor
components $\n R^{i}_{\; jkl}$. The extrinsic curvature curvature
for constant $t$ is given by the expression \beq \label{extr}
K_{ik} = \frac{1}{2N}\left[ N_{i|k} + N_{k|i} - \frac{\partial
g_{jk}}{\partial t} \right], \eeq and the volume element for the
spacetime is $ \sqrt{-\np g} d^n x dt = N \sqrt{\n g} d^n x dt$.
The notation $|k$ represents the covariant derivative along the
spatial direction labelled $k$.

An interesting class of time--dependent solutions would consist in
spacetimes that are interpolating between two different vacua. The
boundary (at conformal infinity) of these spacetimes would then be
disconnected. This is the time--dependent equivalent of static
(Lorentzian or Euclidean) wormholes. Provided the matter sources
generating these time--dependent geometries satisfy certain
(reasonnable) energy conditions, it is clear that such spacetimes
cannot exist (this is detailed in section~\ref{bounces}). We will
see that vacua interpolating spacetimes can be realized only if
something prevents the formation of singular points. From the
point of view of the gravitational field such a singularity
resolution mechanism could, for example, take the form of a bounce
or something that stabilizes the scale of the spatial sections to
finite size. Another possibility is that the volume of the Cauchy
surfaces could be eternally contracting (or expanding) in such a
way that the spacetime does not develop singular points and
therefore remains geodesically complete (non--singular big--bang
or big--crunch). We find explicit examples of these phenomena in
section~\ref{spacelikeb} while in section~\ref{bounces} we consider in
detail the dynamics of bounces. As will become clear shortly a
minimal requirement for these mechanisms of singularity resolution
to take effect it that positive acceleration be allowed by the
matter sources in the system.

Finding analytic solutions of the general form (\ref{general}) is
very difficult. However to investigate the possibility of getting
bouncing geometries and spacetimes with positive acceleration the
first step consists in studying the local dynamics (global aspects
are considered in section~\ref{singular}). We therefore use
Gaussian normal coordinates to write down the metric locally in
the form \beq \label{localmetric} ds^2 = -dn^2 + \n g_{ij} dx^i
dx^j , \eeq where $n$ is a timelike coordinate normal to the
hypersurfaces with spatial metric $\n g_{ij}$. An example we
consider in further detail later is that of an anisotropic
foliation in the form of a product geometry. This assumes
that the region close to the bounce is of the
form \beq \label{localform} ds^2 = -dn^2 + \p
g_{\hat{i}\hat{j}}dx^{\hat{i}} dx^{\hat{j}} + \nmp g_{ab} dx^a
dx^b, \eeq where $\hat{i},\hat{j}=1,\, ... \, , (n-p)$ and
$a,b=(n-p+1),\, ... \, , n$. This is relevant to the study of
gravitational fields generated by unstables D--branes
\cite{strominger1,gutperle,myerssbrane,buchel1,peet,alex} which is
expected if we follow our intuition gained studying regular static
branes \cite{pbranes}. In this case the spacetime close to the core of the
object is of the form eq.~(\ref{localform}) with $n$ a spatial
coordinate and $\n g_{\hat{i}\hat{j}}$ a Lorentzian metric.

We now proceed and write down the Einstein equations associated
with the metric (\ref{localmetric}). We begin by using the
Gauss--Codazzi equation (see, \eg, ref.~\cite{mtw}) to express the
relevant Riemann tensor components in terms of $n$--dimensional
quantities, \ie, the intrinsic and the extrinsic curvatures of the
spatial hypersurfaces, \beq \label{riem1} \np R^{m}_{\;\;ijk} = \n
R^{m}_{\;\;ijk} + \frac{1}{n^2} \left[ K_{ij}K_{\;\;k}^{m} -
K_{ik}K_{\;\;j}^{m} \right], \eeq \beq \label{riem2}\np
R^{n}_{\;\;ijk} = -\frac{1}{n^2}\left[ K_{ij|k}-K_{ik|j} \right],
\eeq \beq \label{riem3} \np R^{n}_{\;\;ink} =
\frac{1}{n^2}\left[\frac{\partial K_{ik}}{\partial n} +
K_{im}K^{m}_{\;\;k} \right], \eeq where $n$ is the timelike vector
normal to the hypersurfaces.\footnote{The spacetimes of interest
are time orientable which implies that we can define a smooth
non--vanishing timelike vector everywhere.} It is then
straightforward to write down the Ricci tensor components using
eqs.~(\ref{riem1})--(\ref{riem3}), \beq \label{insys} \np R_{ik} =
\n R_{ik} + \frac{1}{n^2}\left[ \frac{\partial K_{ik}}{\partial n}
+ 2 \left( K^2 \right)_{ik} - K_{ik} \tr K \right], \eeq \beq \np
R_{ni} = \frac{g_{nn}}{n^2} \left[ -K_{ik}^{|k} + \left(\tr
K\right)_{|i} \right], \eeq \beq \np R_{nn} = \frac{g_{nn}}{n^2}
\left[ \frac{\partial \tr K}{\partial n} - \tr K^{2} \right], \eeq
where $g_{nn}=-1$ and $n^2=-1$. Finally the Ricci
scalar is found to be \beq \np R = \n R + \frac{1}{n^2} \left[ 2
\frac{\partial \tr K} {\partial n} - \tr K^2 - \left(\tr K
\right)^2 \right]. \eeq

We are now in a position to write down the Einstein equations,
\beq  \np G_{\mu\nu} = \np R_{\mu\nu} - \frac{1}{2}g_{\mu\nu} \np R = 8\pi
G_{\scriptscriptstyle N} T_{\mu\nu} \eeq ($\mu,\nu=0,1,\, ... \, , n$),
in terms of the intrinsic and extrinsic curvature of the Cauchy surfaces,
\begin{eqnarray} 8\pi G_{\scriptscriptstyle N} T_{ik} = \np G_{ik} =
\n G_{ik} + \\ \nonumber
\frac{1}{n^2}\left[ \frac{\partial K_{ik}}{\partial n} - K_{ik}
\tr K + 2\left(K^{2}\right)_{ik} - g_{ik}\frac{\partial \tr
K}{\partial n} + \frac{1}{2} g_{ik} \tr K^{2} +
\frac{1}{2}g_{ik}\left(\tr K\right)^{2} \right], \end{eqnarray} \beq 8\pi
G_{\scriptscriptstyle N} T_{ni} = \np G_{ni} = \frac{g_{nn}}{n^2}
\left[ -K_{ik}^{|k} + \left(\tr K\right)_{|i} \right], \eeq \beq
\label{fconstraint} 8\pi G_{\scriptscriptstyle N} T_{nn} = \np
G_{nn} = -\frac{1}{2}g_{nn} \n R + \frac{g_{nn}}{n^2} \left[
-\frac{1}{2}\tr K^{2} + \frac{1}{2} \left( \tr K \right)^{2}
\right]. \eeq The quantity $G_{\scriptscriptstyle N}$ is the
$(n+1)$--dimensional Newton constant. It is important to recall
that strictly speaking the equations of motion we derived are valid only locally
when the metric is written in Gaussian normal coordinates.
However beginning with section~\ref{hubble} we will consider, for simplicity,
homogeneous (but anisotropic) global geometries of the form (\ref{localmetric}).

Let us be more precise with respect to the correct procedure for finding
solutions associated with the equations derived above. We want to
bring these down to a system of (at most) $2\times n(n+1)/2$ first
order differential equations in $\n g_{ik}$ and $K_{ik}$. Using
eqs.~(\ref{extr}) and (\ref{localmetric}) we can write \beq
\label{eq1} K_{ik} = -\frac{1}{2}\frac{\partial g_{ik}}{\partial
n}. \eeq From eq.~(\ref{insys}) we have \beq \label{eq2}
\frac{\partial K_{ik}}{\partial n} = \n R_{ik} - \left[ 2 \left(
K^2 \right)_{ik} - K_{ik} \tr K \right] - 8\pi G_{\scriptscriptstyle
N} \left[ T_{ik} - \frac{g_{ik}}{n-1} T \right]. \eeq
Eqs.~(\ref{eq1}) and (\ref{eq2}) act as evolution equations
respectively for $\n g_{ik}$ and $K_{ik}$. This system of $n(n+1)$
first order differential equations must be supplemented
by the evolution equations associated with the matter fields.
It may also be verified that if eq.~(\ref{fconstraint}) is
satisfied on some initial constant $t$ hypersurface then it will hold
at all times by virtue of the equations of motion. Therefore
eq.~(\ref{fconstraint}) acts as a constraint on the initial values
of $\n g_{ik}$, $K_{ik}$ and the first derivatives of the matter
fields acting as sources. For example in order to find
vacuum solutions to the Einstein equations we must set
$T_{\mu\nu}$ to zero and solve eqs.~(\ref{eq1}) and (\ref{eq2})
while making sure the initial conditions satisfy the constraint
eq.~(\ref{fconstraint}).

\subsection{Attraction and energy conditions}
\label{energy}

So far we have considered the equations governing the spacetime
curvature given arbitrary sources. In this section we review the
effects that the resulting curvature will have on the behavior of
geodesics of physical interest. This will be useful in
section~\ref{singular} where we review singularity theorems
relevant to cosmology in ($n+1$) dimensions.

We consider a smooth congruence of timelike geodesics parametrized
with the affine parameter $\tau$.\footnote{Our approach in this subsection is simply
to generalize results from chapter~8 of ref.~\cite{wald} to
($n+1$) dimensions.} The associated vector field $\xi^{\mu}$
is normalized such that $\xi^{2}=-1$. We introduce the spatial
vector field $B_{\mu\nu}= \nabla_{\nu}\xi_{\mu}$, the symmetric
part of which is related to the extrinsic
curvature through $K_{ij} = - B_{(ij)}$.
Then we consider a smooth one--parameter sub--family
$\gamma_{s}(\tau)$ of geodesics on the congruence and we let
$\eta^{\mu}$ represent an infinitesimal spatial displacement from
$\gamma_{0}$ to a nearby geodesic within the sub--family. It is easy to
see that $\xi^{\mu}\nabla_{\mu}\eta^{\nu} =
B^{\nu}_{\mu}\eta^{\mu}$, which implies that $B$ is a
linear map measuring how much an observer on $\gamma_{0}$
would see the nearby geodesics being stretched and rotated.
Introducing the spatial metric $h_{\mu\nu} = g_{\mu\nu} +
\xi_{\mu}\xi_{\nu}$, we decompose $B$ into
symmetric--traceless, anti--symmetric and
scalar parts, \beq B_{\mu\nu} = \sigma_{\mu\nu} +
\omega_{\mu\nu} + \frac{1}{n}\theta h_{\mu\nu}. \eeq These different
components are given by \beq \sigma_{\mu\nu} = B_{(\mu\nu)} -
\frac{1}{n} \theta h_{\mu\nu}, \;\;\; \omega_{\mu\nu} =
B_{[\mu\nu]}, \;\;\; \theta = B^{\mu\nu}h_{\mu\nu} = B, \eeq
which respectively correspond to the shear, the twist and the
expansion of the congruence. The equations governing the time
evolution of this tensor are shown to be \beq \label{ray} \xi^{\mu}
\nabla_{\nu}B_{\lambda\rho} = -B^{\mu}_{\rho}B_{\lambda\mu} +
R_{\mu\rho\lambda}^{\kappa}\xi^{\mu}\xi_{\kappa}. \eeq For us the
most relevant component is the trace of eq.~(\ref{ray}), \beq
\label{ray2} \xi^{\mu}\nabla_{\mu} \theta = \frac{d\theta}{d\tau} =
-\frac{1}{n}\theta^{2} -\sigma_{\mu\nu}\sigma^{\mu\nu} +
\omega_{\mu\nu}\omega^{\mu\nu} - R_{\mu\nu}\xi^{\mu}\xi^{\nu}.
\eeq This is the Raychaudhuri equation
which describes the rate of expansion of nearby geodesics in a
congruence.\footnote{The symmetric trace--free part of
eq.~(\ref{ray}) governs the dynamics of $\sigma_{\mu\nu}$ while
the anti--symmetric components would give information about
$\omega_{\mu\nu}$.} From now on we set $\omega_{\mu\nu}=0$, \ie , we consider
hypersurface orthogonal spacetimes only. The analysis presented for
timelike geodesics can be repeated in the case of null geodesics
by introducing the tensor $\hat{B}_{\mu\nu} =
\nabla_{\nu}k_{\mu}$, where $k^{\mu}$ is a null vector.
It is straightforward to derive the equation
governing the expansion of null geodesics $\gamma_{s}(\lambda)$ in
a congruence, \beq \frac{d\theta}{d\lambda} =
-\frac{1}{n-1}\theta^{2}
-\hat{\sigma}_{\mu\nu}\hat{\sigma}^{\mu\nu} +
\hat{\omega}_{\mu\nu}\hat{\omega}^{\mu\nu} -
R_{\mu\nu}k^{\mu}k^{\nu}. \eeq

Our point of view (essentially following that presented in
ref.~\cite{wald}) is that energy conditions on matter sources
impose that gravity is attractive which is equivalent to requiring
that \beq \frac{d\theta}{d\tau} \geq 0 \;\;\; {\rm and} \;\;\;
\frac{d\theta}{d\lambda} \geq 0 \eeq everywhere. In the case of
null geodesics this corresponds to requiring that all sources
satisfy \beq T_{\mu\nu}k^{\mu}k^{\nu} \geq 0. \eeq This is called
the null energy condition ({\small NEC}) and is enough to insure,
for example, that converging null rays will never re--expand.

It is believed that any physically reasonable system is associated
with a stress--energy tensor that can be diagonalized.\footnote{A
notable exception being the example of a null fluid (see
ref.~\cite{wald}).} It will therefore be useful to consider
systems of the form \beq \label{stresst} T_{\mu\nu} = (\rho,
g_{i_{1}j_{1}}p_{1}, \, ... \, , g_{i_{N}j_{N}}p_{N}), \eeq where
$\rho$ is the energy density and the $p_{i}$'s are normal pressures
($i=1,\,...\, , N$). Then the {\small NEC} corresponds to \beq
\label{ec} \rho + p_{i} \geq 0 \;\;\; \forall \, i,  \eeq with no
constraint on the sign of $\rho$. The quantity
$T_{\mu\nu}\xi^{\mu}\xi^{\nu}$ physically represents the energy
density of matter as measured by an observer whose
$(n+1)$--velocity is $\xi^{\mu}$. The weak energy condition
({\small WEC}) corresponds to the requirement that
$T_{\mu\nu}\xi^{\mu}\xi^{\nu}\geq 0$ which supersedes the {\small
NEC} by constraining the energy density to be positive ($\rho \geq
0$). Now the strong energy condition ({\small SEC}) corresponds to
the statement that \beq R_{\mu\nu}\xi^{\mu}\xi^{\nu} = 8\pi
G_{\scriptscriptstyle N} \left[ T_{\mu\nu}\xi^{\mu}\xi^{\nu} -
\frac{1}{n-1} ({\rm Tr}\;T) \xi^{\mu}\xi_{\mu} \right] \geq 0,
\eeq which implies that eq.~(\ref{ec}) is satisfied, $\rho\geq 0$
and \beq \left( \sum_{i=1}^{N} n_{i} - 2 \right)\rho +
\sum_{i=1}^{N} n_{i} p_{i} \geq 0, \eeq where $n_{i}$ is the
number of spatial directions with normal pressure $p_{i}$. For a
perfect fluid in four dimensions the latter inequality becomes the
familiar $\rho+3p\geq 0$. This condition plays a crucial role in the
derivation of Hawking--Penrose singularity theorems \cite{hawking}
relevant for cosmology (see section~\ref{singular} for a
review). Clearly de~Sitter space does not respect this energy
condition as manifested by the bounce present in its global
representation (see, \eg, ref.~\cite{volovich}).

The energy conditions presented so far are not fundamental since
they have not been derived from first principle in any theory
containing gravity. However it is interesting (and important for
the problems of interest here) to note that any excitation of the
massless bosonic closed string fields in the ten-- or
eleven--dimensional supergravity theories respect the {\small
SEC}. A notable exception is the 9--form field in massive Type IIA
supergravity (see section~\ref{spacelikeb} for an application).
Also there are non--perturbative objects in string theory which
violate the {\small SEC}. In particular, $p$--branes with $p\geq
8$ are repulsive with a stress--tensor of the form \beq \label{st}
T_{\mu\nu} = T_{p}(1,-1,\, ... \, , -1)\delta({\bf y}), \eeq where
${\bf y}$ represents the spatial directions transverse to the
brane. In fact in the Newtonian limit the gravitational field
sourced by these objects corresponds to\beq \label{classical}
\nabla^{2}\phi = 4\pi G_{\scriptscriptstyle N} \left[ (n-2)\rho +
p T_{i}^{\; i}\right], \eeq where no sum is implied and $\phi$ is
the usual classical gravitational potential. Using eq.~(\ref{st})
we find that the right--hand--side (RHS) of eq.~(\ref{classical})
is negative (repulsive gravity) for $p>(n-2)$. In ten--dimensional
supergravity this means that 8--branes (domain walls) and the
space--filling 9--branes are repulsive. This is a general
statement that applies to any co--dimension one or zero tensile
object in a gravitational theory with any number of dimensions.
Orientifold planes (negative tension objects) also violate the
{\small SEC} for $p\leq 7$. The repulsive objects described here
are static so we should naively not expect them to act as sources
for time--dependent geometries (however see section~\ref{spacelikeb}).

The natural generalization of D--branes to time--dependent
phenomena is to consider the unstable D--branes present in the
non--perturbative spectrum \cite{sen1}. In this case the
instability is caused by the presence of an open string
worldvolume tachyon in the perturbative spectrum. As shown in
ref.~\cite{sen2} the stress--energy tensor associated with an
homogeneous open string tachyon on an unstable D$p$--brane
is\footnote{These expressions are derived from a worldsheet
approach on which the tachyon takes the form of the
marginal boundary deformation: $\tilde{T}(t)=\lambda \cosh (t/\sqrt{2})$.} \beq
\label{tachyonform1} \rho =
\frac{T_{p}}{2}\left(\cos(2\lambda\pi+1)\right), \eeq \beq
\label{tachyonform2} p= -T_{p}\left[
\frac{1}{1+e^{\sqrt{2}t}\sin^{2}(\lambda\pi)} +
\frac{1}{1+e^{-\sqrt{2}t}\sin^{2}(\lambda\pi)} - 1 \right],\eeq
where $T_{p}$ is the brane tension. As
expected if we consider the latter as a gravitational source it
violates the {\small SEC} in ten dimensions for
$p\geq 7$ just like the static D$p$--branes.

Lastly we consider the dominant energy condition ({\small DEC})
which requires that ($-T^{\mu}_{\nu}\xi^{\nu}$) is a
future--directed timelike or null vector. For an observer with
($n+1$)--velocity $\xi^{\mu}$ this vector measures the
energy--momentum current of matter she observes. The {\small DEC}
can be interpreted as a constraint imposing that the speed of
energy flow of matter is less than the speed of light. It clearly
makes sense physically for matter to possess this characteristic.
If the sources are in the form of a perfect fluid, then the
{\small DEC} implies that \beq \rho \geq |p_{i}| \;\;\; \forall \,
i. \eeq We also note that the {\small WEC} is implied by this last
condition.

\subsection{Cosmological singularity theorems}
\label{singular}

Before we discuss the physics of bounces and cosmological
acceleration, we briefly review some formal notions about
singularities. If the {\small SEC} is satisfied then
eq.~(\ref{ray}) implies the inequality \beq \frac{d
\theta^{-1}}{d\tau} \geq \frac{1}{n}, \eeq which can be integrated
to give \beq \theta^{-1}(\tau) \geq  \theta_{0}^{-1} +
\frac{\tau}{n}, \eeq where $\theta_{0}$ is the expansion rate on
an arbitrary spatial hypersurface. If $\theta$ is negative (rays
are converging) then eq.~(\ref{ray}) implies that $\theta$ becomes
infinite within $\Delta\tau=(n-1)/\theta_{0}$. This is a
pathological behavior but to conclude that a spacetime is singular
it is not enough to show that it contains conjugate points. For a
spacetime to be called singular it must also contain maximal
length curves.

In the rest of this paper we will be referring to two
important singularity theorems.\footnote{The proofs for all singularity
theorems relevant for ($3+1$)--dimensional physics can be found
in refs.~\cite{hawking,wald}. Their generalization to ($n+1$)--dimensional gravity
is straightforward and implies that the theorems hold without modification for $n\neq 3$. The case
$n=1$ is special.} If the conditions stipulated in
these theorems are satisfied then the corresponding spacetime is
necessarily singular in the sense of timelike or null geodesic
incompleteness. An important point is that the singulartiy
theorems have nothing to say about the nature of the singularity.
However they would appear to predict a breakdown of general relativity.
For the first singularity theorem we paraphrase theorem~9.5.1
from ref.~\cite{wald} where the proof can be found:

\begin{quotation}
\noindent {\bf Theorem I:} Suppose $({\cal M},g_{\mu\nu})$ is a
globally hyperbolic spacetime with
$R_{\mu\nu}\xi^{\mu}\xi^{\nu}\geq 0$ for $\xi^{\mu}$ a timelike
vector. If there exists a smooth spacelike Cauchy surface for
which the trace of the extrinsic curvature is strictly negative
($-\theta=\tr K < 0$) everywhere, all past directed timelike
geodesics are incomplete.
\end{quotation}

\noindent This is a powerful statement guaranteeing that if a
globally hyperbolic cosmological spacetime is everywhere expanding
at a finite rate it must have begun in a singular state a finite
time ago. The same conclusion must be reached with respect to the
future of a spacetime that is everywhere contracting. This theorem
has two weaknesses. Firstly it requires global hyperbolicity and,
secondly, it says nothing about the role played by inhomogeneities
during gravitational collapse. We therefore review another
singularity theorem, due to Penrose and Hawking\cite{hawking},
which remedies this:

\begin{quotation}
\noindent {\bf Theorem II:} Suppose a spacetime $({\cal
M},g_{\mu\nu})$ satisfies $R_{\mu\nu}v^{\mu}v^{\nu}\geq 0$ for all
timelike and null vectors $v^{\mu}$ as well as the timelike and
null generic energy conditions. Then if the spacetime is a closed
cosmology or there exists a point $p$ such that the expansion of
the future (or past) directed null geodesics emanating from $p$
becomes negative along each geodesic in this congruence, then
$({\cal M},g_{\mu\nu})$ must contain at least one incomplete
timelike or null geodesic.
\end{quotation}

\noindent The most important aspect of this theorem for us is that
it eliminates the assumption that the spacetime is expanding
everywhere on the spatial hypersurfaces. In principle this allows
one to treat cases where collapsing geometries are inhomogeneous.
The new ingredient in this formulation is that the generic
timelike and null energy condition must be satisfied. This will be
true if each timelike (null) geodesic in (${\cal M},g_{\mu\nu}$)
possesses at least one point where
$R_{\mu\nu\rho\lambda}\xi^{\mu}\xi^{\lambda}\neq 0$
($R_{\mu\nu\rho\lambda}k^{\mu}k^{\lambda}\neq 0$). It would be
surprising to find interesting cosmological spacetimes that do not
satisfy these generic conditions. For example, for a homogeneous
and isotropic FLRW cosmology the timelike condition implies that
$\ddot{a}\neq 0$ somewhere in the spacetime.

\section{Geometrical bounces and singularity resolution}
\label{bounces}

One of our aim is to find manners by which cosmological
singularities are resolved in the context of the effective
theories obtained from string theory. One way this could happen is
if the spacetime is allowed to get out of a phase of contraction
by bouncing. It has long been known that in four--dimensional
conventional Einstein gravity this is prohibited (see, \eg,
ref.~\cite{burgess}) unless the matter supporting the geometry can
violate the {\small SEC}. In this section we review this argument
in all generality and show that it holds for $n\neq 3$ as well.
However we observe that time--dependent bounces should be allowed on
lower--dimensional sub--manifolds when considering dimensional
reductions of higher--dimensional theories for which the {\small
SEC} is {\it not violated}. This can have interesting implications
for cosmology. For example it could mean that a bounce has
occurred in the form of a `non--singular big--bang'
in the past. Of course since the {\small SEC} holds for the
higher--dimensional spacetime the singularity theorems presented
in section~\ref{singular} predict the existence of
singular points. However these could in principle be pushed back
arbitrarily far in the past to times for which no cosmological
data is available. In essence this would imply that in order to
describe theoretically the relevant part of the cosmological
evolution we may not need to worry about quantum gravity.

Clearly all this phenomenon does is allow us to push the
problem of dealing with quantum gravity (at a cosmological level)
back in time. This is based on a general geometrical effect allowing for exotic
phenomena to happen in ($3+1$) dimensions due to the compensating
effect of the dynamics in the transverse dimensions. A related
manifestation of this was recently considered when attempts were
made to explain four--dimensional positive acceleration
in the context of higher--dimensional supergravity theories (see, \eg,
refs.~\cite{townsend1,townsend2,emparan,ulf,today}). More concrete
explanations and examples are provided in later sections.

\subsection{Local dynamics of a bounce}

We now study the physics of gravitational bounces by considering the
local dynamics of the phenomenon. This simply means we are
investigating a spacetime region where the general metric ansatz
(\ref{general}) can be written in the form (\ref{localmetric}) using
Gaussian normal coordinates. The local analysis we present is
complemented in section~\ref{globalaspects} by global considerations
based on the cosmological singulartiy theorems introduced earlier.

We define a bouncing spacetime region as a co--moving volume
element that goes through a minimum in finite time (the parameter
$n$ in this case). The volume element is defined by \beq
\label{vol} V(\delta M) = \int_{\delta M} \sqrt{\n g} d^n x, \eeq
where $\delta M$ denotes an open set of points on a spacelike
hypersurface. It is also legitimate to regard $V(\delta M)$ as
being part of a lower--dimensional manifold in a
higher--dimensional spacetime. This will be relevant to the case
where only some of the spatial dimensions are associated with a
bounce. The integer $n$ will then be understood to mean the number
of bouncing spatial directions. The function $V(\delta M)$ will
reach an extremum when its first order variation vanishes. This
corresponds to the condition \beq \delta V(\delta M) =
\int_{\delta M} \sqrt{\n g} (\tr K) \delta n(x) d^{n}x = 0. \eeq
Consequently a spacelike region associated with a vanishing trace
for the extrinsic curvature corresponds to an extremum for the
volume element. This will be a minimum (a bounce) when the second
order variation of $V(\delta M)$ is positive, \ie, \beq \delta^{2}
V(\delta M) = -\int_{\delta M} \sqrt{\n g} \left[ \left(\tr K
\right)^2 + \frac{\partial \tr K}{\partial n} \right] > 0. \eeq
In summary a spacelike region is said to bounce when $\tr K=0$ and
\beq \frac{\partial \tr K}{\partial n} < 0. \eeq This conclusion
is valid locally, \ie, when the metric can be written like
eq.~(\ref{localmetric}) and for a small patch of a given
hypersurface. This is easily generalized to the case where an
entire hypersurface, obtained by integrating over all volume
elements $\delta M$, is bouncing.

\subsection{Solitary bounces and energy conditions}
\label{solitary}

We have seen earlier that for gravity to be an attractive force
the {\small NEC} and the {\small SEC} need to be satisfied. We
also assume that the energy density is positive--definite ({\small
WEC}) for the cases of physical interest considered here. These
are characteristics we assume a higher--dimensional effective
theory derived from a fundamental theory such as string theory
must possess. However, as pointed out above, so--called
anti--gravitational effects, \ie, apparent violations of energy
conditions, could happen in lower--dimensional theories obtained
through compactification.

We consider a general spacetime ${\cal M}$ composed of distinct
spatial sub--manifolds \beq \label{product2} {\cal M} = {\cal R}
\times M_{1} \times \, ... \, \times M_{N}, \eeq where ${\cal R}$
represents the timelike variable. For this special case we find
the relations \beq \label{simple1} \tr K = \sum_{a=1}^{N} \tr
K_{a}, \;\;\;\;\;\; \tr K^{2} = \sum_{a=1}^{N} \tr (K_{a})^{2},
\eeq where $K_{a}$ is the extrinsic curvature associated with the
sub--manifold $M_{a}$. The requirement that the energy density be
positive is equivalent to the inequality \beq \label{constraint22}
\rho = \frac{1}{2} \n R +\frac{1}{2} \left[ (\tr K)^{2} - \tr
K^{2} \right] \geq 0. \eeq Clearly for a $N=1$ spacetime to bounce
(at which point $\tr K=0$) the constraint (\ref{constraint22})
becomes \beq \n R \geq \tr K^{2}. \eeq For a ($n+1$)--dimensional
FLRW cosmology this implies that the spatial curvature is positive
because then $\tr K^{2}=0$ at the bounce. This is simply the
well--known result that bounces can only happen for spacetimes
foliated with spherical hypersurfaces.

Let us now consider the case $N>1$ where only one sub--manifold,
say $M_{N}$, goes through a bounce. Then, using
eqs.~(\ref{simple1}) as well as assuming $\tr K_{N}=0=\tr
K_{N}^{2}$, the positive energy condition becomes \beq
\label{theweak} \n R + \sum_{a=1}^{N-1} \left[ (\tr K_{a})^{2} -
\tr K_{a}^{2} \right] \geq 0. \eeq It is clear that for many
spacetimes the second term in eq.~(\ref{theweak}) can be positive
at the bounce. For $N=2$ this term is positive--definite when we
consider a product spacetime of isotropic and homogeneous spaces
(see section~\ref{hubble} for details). This geometric effect
affords us much leeway in getting bounces on spatial
sub--manifolds while maintaining positivity of energy everywhere.
A similar conclusion is reached with respect to having the {\small
WEC} satisfied everywhere.\footnote{The {\small WEC} implies that
$\n R + \tr K^{2} - n (\tr K)^{2} + (n-1)\frac{\partial
K}{\partial n}\geq 0$.} The {\small SEC} requires that
$R_{\mu\nu}\xi^{\mu}\xi^{\nu}\geq 0$ for any timelike vector, \ie
, \beq \label{secfin} \frac{\partial \tr K}{\partial n} - \tr
K^{2} \geq 0. \eeq For $N=1$ this constraint excludes spacetimes
where any portion of a $n$--dimensional hypersurface would bounce.
In the case of a product spacetime of the form (\ref{product2}) if
the sub--manifold $M_{N}$ bounces, the {\small SEC} takes the form
\beq \sum_{a=1}^{N-1}\frac{\partial \tr K_{a}}{\partial n} -
\sum_{a=1}^{N-1} \tr (K_{a})^{2} - \left| \frac{\partial
K_{N}}{\partial n} \right| \geq 0. \eeq This last expression
implies that for a bounce to occur on $M_{N}$ the first term on
the left--hand--side (LHS) must be positive and large enough to
make the whole expression positive. The conclusion is basically
that bounces on sub--manifolds are clearly not excluded even in
theories where the energy conditions are satisfied. This last
statement obviously applies to spacetimes where there is positive
acceleration on $M_{N}$. However in this case one does not need to
worry about the extra constraint that $\tr K_{N}$ must vanish.

\subsection{The compulsory singularity}
\label{globalaspects}

The general system we have in mind is ($n+1$)--dimensional
Einstein gravity coupled to sources originating from the effective
theories describing the low energy dynamics of string theories. We
are also taking the point of view that these higher--dimensional
theories respect the {\small SEC}. As shown above this prevents
the occurrence of phenomena such as positive acceleration and
bounces if the spacetime is of the form ${\cal R}\times
M$,\footnote{$M$ is not necessarily homogeneous and isotropic.}
\ie, when it can be written locally in the form
(\ref{localmetric}). Relevant cosmological backgrounds are such
that, at least in some small interval $\delta n$, the trace of the
intrinsic curvature does not vanish (\ie , $\tr K \neq 0$ on a
subset of Cauchy surfaces). According to Theorem~I presented in
section~\ref{singular}, this implies that singular points must
develop either in the past or the future of this region (depending
on the sign of $\tr K$) if the SEC is satisfied. It might be
relevant to consider cases for which the sign and magnitude of
$\tr K$ vary on a given constant $n$
hypersurface.\footnote{This is important when studying the role of
inhomogeneities.} In this case theorem~II implies that at least
one singular point will form either in the past or the future of
the Cauchy surface under consideration. In other words we expect
that inhomogeneities of supergravity fields cannot prevent the
appearance of singular points in the past of an expanding
phase.\footnote{This conclusion also applies to less conventional
sources such as the $p\leq 7$ tachyon considered in
refs.~\cite{buchel1,peet}.}

We are mostly interested in geometries that can (at least locally)
be written in the form of a product spacetime as in
(\ref{product2}). Again any interesting cosmology will be
associated with at least some Cauchy surfaces where the trace of
the extrinsic curvature does not vanish, \ie, \beq \tr K =
\sum_{a=1}^{N} \tr K_{a} \neq 0 \eeq in some finite interval
$\delta n$. Following the reasoning of the previous paragraph this
implies that singular points will develop either in the future or
the past of this region. Supergravity theories typically respect
the {\small SEC}\footnote{The very few exceptions to this rule are
exploited in section~\ref{spacelikeb}.} which implies that no
regular time--dependent solutions (in the sense of the
cosmological singularity theorems) can be obtained in this
context. Naively we would expect that solutions submitted to the
constraint $\tr K =0$ everywhere will lead to non--singular
cosmologies because the singularity theorems are inapplicable.
This is note the case since the {\small SEC}, which takes the
form (\ref{secfin}), then becomes \beq \tr K^{2} \leq 0. \eeq This
is obviously impossible to satisfy for any non--trivial time--dependent
geometry. This will be illustrated more clearly in
section~\ref{hubble} when a more specific metric ansatz is
considered.

The main conclusion here is that non--singular time--dependent
solutions do not exist in supergravity. This is also true for
spacetimes with phases of positive acceleration and gravitational
bounces. As pointed out in section~\ref{energy} exceptions exist
since there are stringy sources violating the {\small SEC}. This
could include spacetimes supported by a tachyon source associated
with an unstable brane with spatial co--dimension one or zero.
Such gravitational solutions were recently studied in
ref.~\cite{sen10} in the context of bouncing cosmology. The other
exception consists in considering cosmologies supported by a matter content which includes a
space--filling anti--symmetric form--fields.\footnote{If the form
field is the only matter component it essentially plays the role
of a cosmological constant.} Examples related to this are
presented in the remaining sections of this paper.

\subsection{Singular Hubble loops}
\label{hubble}

It is always possible to obtain interesting phenomena such as
positive acceleration and bounces on a sub--manifold in a
higher--dimensional spacetime where the {\small SEC} is {\it not
violated}. The latter condition is violated only effectively from
the point of view say of an observer on the lower--dimensional
manifold. This observation is motivated by the local analysis
presented in section~\ref{solitary} (see ref.~\cite{townsend1} for
the case of positive acceleration). However while the spacetime
might appear to be non--singular with respect to the effective
metric on this sub--manifold, there will always be singular points
in the higher--dimensional realization. This conclusion is reached
by considering global aspects via the cosmological singularity
theorems presented in section~\ref{singular}. Generally phases
with positive acceleration or a gravitational bounce can be used
as a mechanism to avoid the appearance of singular points. Our
point is that this can occur, for instance, on a
($3+1$)--dimensional sub--manifold embedded in a
higher--dimensional theory with sources not violating the {\small
SEC}. This represents a mechanism for which the singularity
resolution on the lower--dimensional spacetime is only an
illusion.

Let us illustrate more concretely this type of behavior. We
consider globally hyperbolic time--dependent geometries composed
of $N$ distinct homogeneous and isotropic spatial sub--manifolds
with different scale factors. The metric ansatz is \beq ds^2 =
-dt^2 + \sum_{i=1}^{N} a_{i}(t)^2 d\Sigma_{n_{i},k_{i}}^{2}, \eeq
where $n_{i}$ is the dimensionality ($n=\sum_{i=1}^{N} n_{i}$) and
$k_{i}$ the spatial curvature associated with each sub--manifold.
The intrinsic curvature of the Cauchy surfaces is given by \beq
\label{spatialsections} \n R = \sum_{i=1}^{N}
\frac{n_{i}(n_{i}-1)k_{i}}{a_{i}^{2}}, \eeq and the average
expansion rate of infinitesimally nearby geodesics is \beq \theta
= -\tr K = \sum_{i=1}^{N} n_{i} H_{i}, \eeq where we introduced
the Hubble factors $H_{i}=\dot{a}_{i}/a_{i}$. Another useful
relation is \beq \label{useful} \tr K^{2} =\sum_{i=1}^{N} n_{i}
H_{i}^{2}, \eeq which allows us to write down \beq \left(\tr K
\right)^{2} -\tr K^{2} = \sum_{i=1}^{N} n_{i}(n_{i}-1)H_{i}^{2} +
\left( {\rm cross-terms} \right), \eeq where the cross--terms can
be either positive or negative depending on the relative sign of
the $H_{i}$'s. Using eq.~(\ref{constraint22}) the energy density
becomes \beq \label{thebig} \rho = \frac{1}{2} \sum_{i=1}^{N}
n_{i}(n_{i}-1)\left( H_{i}^{2} + \frac{k_{i}}{a_{i}^{2}} \right) +
({\rm cross-terms}), \eeq which, for $N=1$, gives the usual \beq
\label{fcons} \rho = \frac{n(n-1)}{2} \left( H^{2} +
\frac{k}{a^{2}} \right). \eeq The {\small SEC} (\ref{secfin}) is
equivalent to the constraint \beq \label{sec1acc} \sum_{i=1}^{N}
n_{i} \frac{\ddot{a}_{i}}{a_{i}} \leq 0, \eeq which cannot be
satisfied for positive acceleration if $N=1$.

\EPSFIGURE[r]{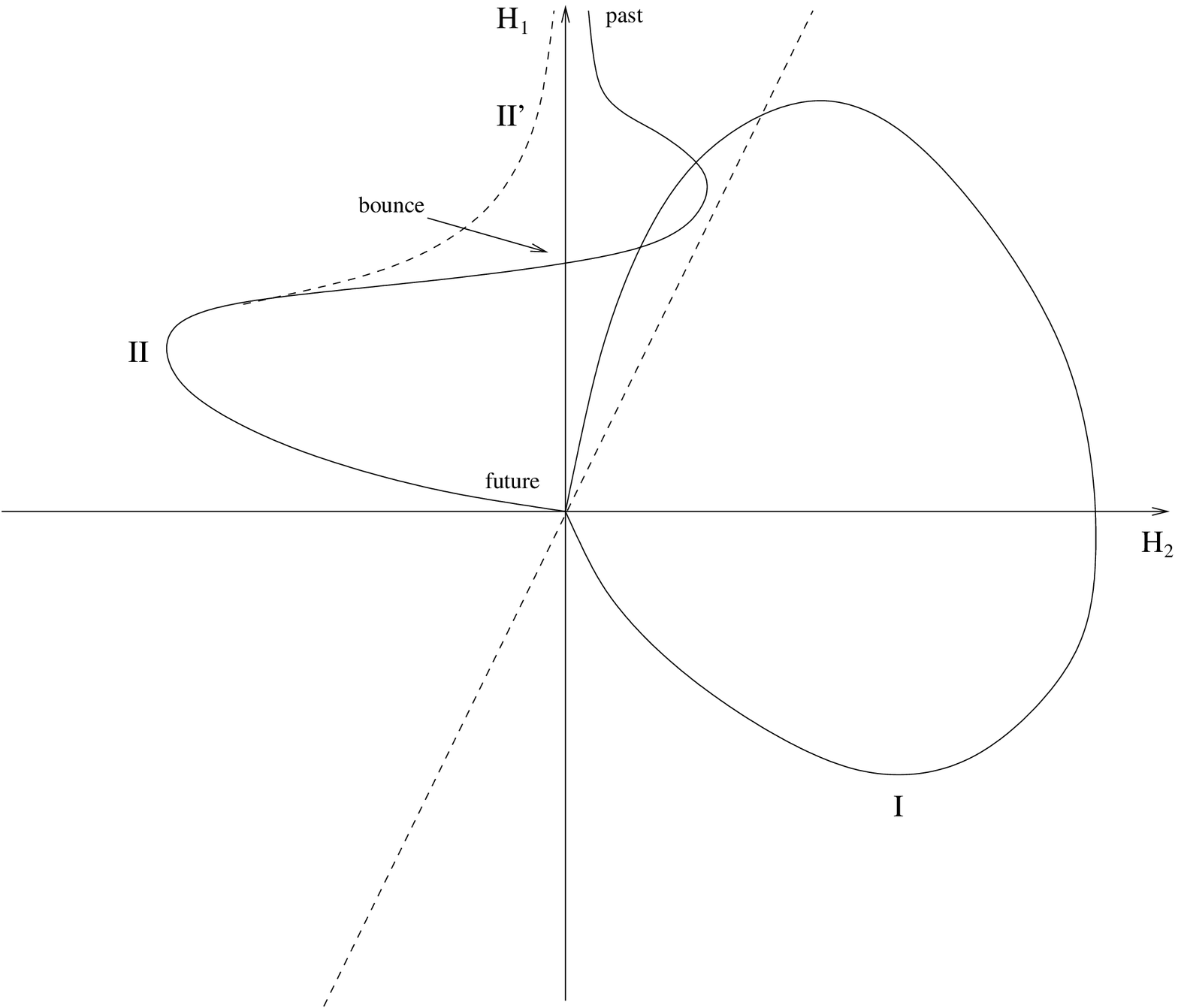,width=80mm}{Typical trajectories in the
($H_{1},H_{2}$) plane associated with $N=2$ cosmological
spacetimes. Curve~I represents a non--singular evolution between
asymptotic regions with $H_{i}\rightarrow 0$ for large $|t|$.
Curve~II shows a singular evolution including a bounce for the
scale factor $a_{2}$ at intermediate time.\label{hubble1}}

We have established earlier that a cosmological spacetime cannot
be associated with a gravitational bounce. However we have shown
that a sub--manifold within a higher--dimensional spacetime can
contain a bouncing phase. The drawback is that while this
sub--space might appear regular, singular points will always
develop as seen from the higher--dimensional point of view. In
what follows we illustrate this generic behavior explicitly for
the case $N=2$.

The assumption we make is that the scale factor $a_{2}$ goes
through a bounce ($\dot{a}_{2}=0$ and $\ddot{a}_{2}>0$). The
expression (\ref{constraint22}) for the density of energy (assumed
positive) at the lower--dimensional bounce becomes \beq
\label{critic} \rho_{c} = \frac{n_{2}(n_{2}-1)k_{2}}{2a_{2}^{2}} +
\frac{n_{1}(n_{1}-1)}{2}\left( H_{1}^{2} + \frac{k_{1}}{a_{1}^{2}}
\right). \eeq The $N=1$ constraint requiring $k=+1$ as a necessary
condition for a bounce is relaxed for $N=2$. Bounces can happen as
long as the contribution to $\rho$ from the transverse directions
is sufficient to make the RHS of eq.~(\ref{critic}) positive. In
particular bounces can in principle occur even for flat
($k_{2}=0$) and hyperbolic ($k_{2}=-1$) foliations.

The volume of the Cauchy surfaces is proportional to the function
\beq \label{volume} V(t) = a(t)^{n} = a_{1}(t)^{n_{1}}
a_{2}(t)^{n_{2}},\eeq where $a(t)$ is an average scale factor. The
second derivative of this expression is \beq \ddot{V} = n\, a^{n}
\left[\frac{\ddot{a}}{a}+(n-1) \left(\frac{\dot{a}}{a}\right)^{2}
\right]. \eeq Using this last expression we find \beq
\label{volume1} n \frac{\ddot{a}}{a} = n_{1}
\frac{\ddot{a}_{1}}{a_{1}} + n_{2} \frac{\ddot{a}_{2}}{a_{2}} -
\frac{n_{1}n_{2}}{n} \left[ H_{1} - H_{2} \right]^{2}.  \eeq The
RHS of eq.~(\ref{volume1}) must be negative since the {\small SEC}
is satisfied in the higher--dimensional theory. If a bounce occurs
on the lower--dimensional manifold with scale factor $a_{2}$, the
constraint (\ref{sec1acc}) becomes \beq \label{condition2} n_{1}
\frac{\ddot{a}_{1}}{a_{1}} \leq - \left| n_{2}
\frac{\ddot{a}_{2}}{a_{2}} \right|. \eeq This implies that during
the lower--dimensional bounce the dynamics of the transverse
dimensions must be such that $\ddot{a}_{1}/a_{1}$ is negative with
a magnitude large enough to compensate for the positivity of the
acceleration associated with $a_{2}$. Of course the constraint
$\ddot{a}/a \leq 0$ must also be satisfied.

To illustrate this we consider a pictorial approach. All $N=2$
cosmological evolutions can be represented by a parametric curve
in the $(H_{1},H_{2})$ plane. Non--singular asymptotically flat
(or, more precisely, asymptotically FLRW) solutions would be
associated with trajectories asymptoting (\ie, in the infinite
past and future) to the point $(H_{1}=0,H_{2}=0)$. The manner by
which these asymptotic regions are reached is crucial information
with respect to the global features of the spacetime. In
particular the `speed' with which the attractor points are
attained will be determined by the dominant matter component at
early and late time. Consequently all non--singular cosmological
evolutions should be represented by closed loops in the
two--dimensional $(H_{1},H_{2})$ plane (see, for example, curve~I
on figure~\ref{hubble1}).\footnote{The generalization to
spacetimes with more sub--spaces is straightforward.} However we
have seen that in the context of higher--dimensional gravitational
theories respecting the {\small SEC} the allowed trajectories
cannot be closed based on the application of cosmological
singularity theorems. In order to obtain closed trajectories new
ingredients violating the {\small SEC} should be introduced. In
section~\ref{spacelikeb} we provide examples of this for
asymptotically de~Sitter (dS) spacetimes. Of course in this case
the location of the endpoints of the $(H_{1},H_{2})$ trajectories
is changed.

A more realistic trajectory is represented by curve~II on
figure~\ref{hubble1}. It corresponds to a geometry evolving out of
a big--bang in the infinite past (when $H_{1}\rightarrow +\infty$
and $H_{2}\rightarrow 0$), passing through a phase where $a_{2}$
bounces ($H_{2}=0$ and $H_{1}$ is finite) and finally evolving
toward the ($H_{1}=0,H_{2}=0$) point in the infinite
future.\footnote{The time--reversed evolution would correspond to
a spacetime evolving into a big--crunch singularity.} The
effective gravity seen on the manifold with scale factor $a_{2}$
appears non--singular (\eg,
$\lim_{t\rightarrow\pm\infty}H_{2}=0$). However the Hubble factor
$H_{1}$ blows up at $t=-\infty$ which suggests that there is a
curvature singularity in the past. Such a spacetimes would be an
example of the mirage singularity resolution described earlier.

Physically allowed trajectories on the $(H_{1},H_{2})$ plane must
satisfy other constraints such as positivity of the energy
density, the {\small SEC} and the {\small NEC}. The line $n_{1}
H_{1}+n_{2}H_{2}=0$ (dashed line on figure~\ref{hubble1}) is
special because it corresponds to an extremum of the volume
function $V(t)$. Of course it cannot be traversed  if the extremum
corresponds to a minimum, \ie, a bounce. A different example is
depicted by curve~II' on figure~\ref{hubble1}. In this case there
is no bounce associated with $a_{2}$ and a curvature singularity
is still present in the future. As for curve~II the effective
geometry associated with $a_{2}$ is non--singular. Because there
is no bounce the corresponding effective spacetime will be regular
and forever expanding. This situation is
allowed because the {\small SEC} is effectively violated and the
singularity theorems lose their predictive power.

\section{Spacetimes with positive acceleration}
\label{sugra}

Simple compactifications of supergravity theories
lead to lower--dimensional effective actions where moduli fields
such as the dilaton acquire potentials. These
are typically of the form \beq \label{thepot} V(\phi) = \Lambda e^{-\alpha\phi}, \eeq
where the scale $\Lambda$ is set by the magnitude of fluxes and/or the internal curvature.
In section~\ref{applications} we consider explicit examples. The constant
$\alpha$ is also determined by the ingredients present in
the compactification scheme. We will see that its value is critical
in the determination of whether or not non--singular cosmological solutions exist.
Potentials of the form (\ref{thepot}) are of interest to us because for $\Lambda>0$
the corresponding scalar field can violate the {\small SEC}.

In this section we provide a study of spacetimes supported by
homogeneous scalar fields with potentials of the form
(\ref{thepot}). In section~\ref{sec1} we begin by reviewing FLRW
cosmology in ($m+1$)--dimensions and explain some relevant aspects
associated with spacetimes for which the {\small SEC} is violated.
Then in section~\ref{sec2} we present analytic solutions for
spacetimes with flat foliations and study their qualitative
features. In section~\ref{sec3} we study the corresponding
spacetimes with positive spatial curvature and, in particular, we
consider geometries containing a bounce. All non--singular
spacetimes we consider in this section are asymptotically FLRW
with positive acceleration. In section~\ref{applications} we will
consider uplifting the solutions described in this section to
solutions of ten-- and eleven--dimensional supergravity.

\subsection{Bounces and FLRW cosmology}
\label{sec1}

Before proceeding with examples we make general remarks about
conventional homogeneous and isotropic ($m+1$)--dimensional FLRW
cosmology. These comments are not novel in any way but will be
useful in the remaining sections of this paper.

The invariance of gravitational actions under diffeomorphisms
implies that the stress--energy tensor is covariantly conserved,
\ie, $\nabla^{\mu}T_{\mu\nu} = 0$. For a metric of the form \beq
ds^{2} = -dt^{2} + a(t)^{2} d\Sigma_{m,k}^{2}, \eeq the continuity
equation implies \beq \label{contin} \dot{\rho} = -m
\frac{\dot{a}}{a}\left( \rho + p \right). \eeq By inspection of
eq.~(\ref{contin}) we note that gravitational sources that are not
respecting the {\small WEC} lead to pathologies. In fact if
$\rho+p < 0$ an expanding universe is associated with an
increasing density of energy. For $\rho = -p$ the energy density
is constant which corresponds to dS space.

We assume that the sources obey the simple equation of state
$p=w(t)\rho$. Solving eq.~(\ref{contin}) we find \beq
\label{nouse} \ln \rho = - m \int dt (1+w) \frac{d \ln a}{dt}.
\eeq Clearly $w(t)$ is not known beforehand but it can be useful
to assume it to be a constant. This can be an acceptable
approximation when studying different phases of a given cosmology.
The solutions to eq.~(\ref{nouse}) are then of the form\beq
\label{state} \rho = C a^{-m(1+w)}, \eeq where $C$ is a constant.
Frequently encountered cases include pressureless
non--relativistic dust ($\rho \sim a^{-m}$, $w=0$), radiation
($\rho \sim a^{-(m+1)}$, $w=m^{-1}$) and a cosmological constant
($w=-1$). Other types of sources that are more stringy include,
for example, the tachyon matter ($w =0$) \cite{sen2} and a gas of
$\tilde{p}$--branes for which the equation of state is
\cite{easson} \beq w = \frac{(\tilde{p}+1)v^2 - \tilde{p}}{m+1},
\eeq where $v$ is the magnitude of the average velocity associated
with these spatially extended objects. It is interesting to note
that a gas of $\tilde{p}$--branes violates the {\small SEC} for
$\tilde{p}\geq 8$ as was found for conventional static and
unstable D$\tilde{p}$--branes in section~\ref{energy}.

In order to determine which component among the sources will
dominate at different moments of the evolution it is useful to
consider the ratio \beq \frac{\rho_{w_{1}}}{\rho_{w_{2}}} \sim
\frac{1}{a^{m(w_{1}-w_{2})}}. \eeq The integration constant in
eq.~(\ref{state}) is important so studying such ratios gives us
only a crude understanding of the system of interest. Clearly when
$w_{1}>w_{2}$ the component associated with $w_{2}$ dominates for
large values of the scale factor (large spatial volumes) and,
conversely, the component $w_{1}$ dominates for small spatial
volumes. For example, the ratio for pressureless dust and
radiation is $ \rho_{r}/\rho_{m} \sim 1/a$, which leads to the
well--known result that the late time (large $a$) evolution of the
universe is matter--dominated. The density ratio of cosmological
constant to any kind of matter with equation of state $w$ is
$\rho_{\Lambda}/\rho_{w}\sim a^{m(1+w)}$. For matter respecting
the {\small SEC} it is clear that for early time dynamics (small
$a$) a cosmological constant term tends to be overwhelmed. At late
times (large $a$) however the cosmological constant always dominates.

The point of view here is that the ($m+1$)--dimensional theory of
gravity we are studying is derived from a higher--dimensional
theory such as supergravity. This is why we consider a matter
content which can in principle violate the {\small SEC}, \ie, it
can induce periods of positive acceleration. A certain combination
of the field equations gives \beq \label{field1}
\frac{\ddot{a}}{a} = -\frac{8\pi G_{\scriptscriptstyle
N}}{m-1}\left[(m-2)\rho + m p \right]. \eeq Clearly the sign of
the RHS in eq.~(\ref{field1}) determines whether the acceleration
is positive or negative. As pointed out earlier the {\small SEC}
requires that $(m-2)\rho + m p\geq 0$ which implies that sources
satisfying this can only support negative acceleration. This also
implies that isotropic and homogeneous $(m+1)$--dimensional
gravitational backgrounds sourced by matter respecting the {\small
SEC} cannot bounce. This is clearly not the case if $-1 \leq w<
\left(\frac{2}{n}-1\right)$, \ie, when $w$ is such that the
{\small SEC} is violated.\footnote{The WEC is violated for
$w<-1$.} This includes dS space which is in fact a bouncing
spacetime when written in global coordinates.

The Friedmann constraint (\ref{fcons}) can be written in the form
\beq \label{field2} \frac{1}{2}\dot{a}^2 - \frac{8\pi
G_{\scriptscriptstyle N}}{m(m-1)} \left(\rho a^{2}\right) =
-\frac{k}{2}. \eeq This is equivalent to the first order equation
governing the classical dynamics of a point particle if we replace
$a$ with the spatial displacement $x$. The conserved energy is
then $-k/2$ and the potential function is given by \beq V(a) =
-\frac{8\pi G_{\scriptscriptstyle N}}{m(m-1)}\left(e^{- m\int dt
(1+w) \frac{d \ln a}{dt}} a^{2}\right). \eeq As pointed out above
this is a useful analogy only if the equation of state is not
time--dependent in which case we get \beq \label{expo} V(a) =
-\frac{8\pi G_{\scriptscriptstyle N}}{m(m-1)}
\frac{C}{a^{m(1+w)-2}}, \eeq where $C>0$ ($C<0$) for positive
(negative) energy density. For $k=0$ the solution to
eq.~(\ref{field2}) is of the form $a\sim t^{\frac{2}{n(1+w)}}$ for
$w\neq 1$ and for $w=-1$ we get the usual dS exponential.

As pointed out earlier we are interested in spacetimes that are
bouncing or, at least, include phases of positive acceleration.
Bouncing spacetimes have co--moving volumes evolving in such a way
as to connect two different asymptotic vacua with large and
possibly forever expanding spatial volumes. As is clear from the
Friedmann constraint, matter sources satisfying the {\small WEC}
can lead to a bounce only if the spatial curvature is positive
($k=+1$). The sources that support these gravitational backgrounds
must therefore dominate the spatial curvature at late and early
time ($t\rightarrow\pm\infty$) in order to prevent the apparition
of cosmological singularities. This characteristic is also
required of a realistic cosmological model if it is expected to
conform with the observation suggesting the spatial curvature
is currently very small \cite{supernovae}. The latter condition
will be satisfied for realistic cosmological models predicting that \beq
\frac{k/a^2}{8\pi G_{\scriptscriptstyle N} \rho} \eeq is presently
small. In principle this implies our universe could have
negative spatial curvature as well.
\EPSFIGURE[r]{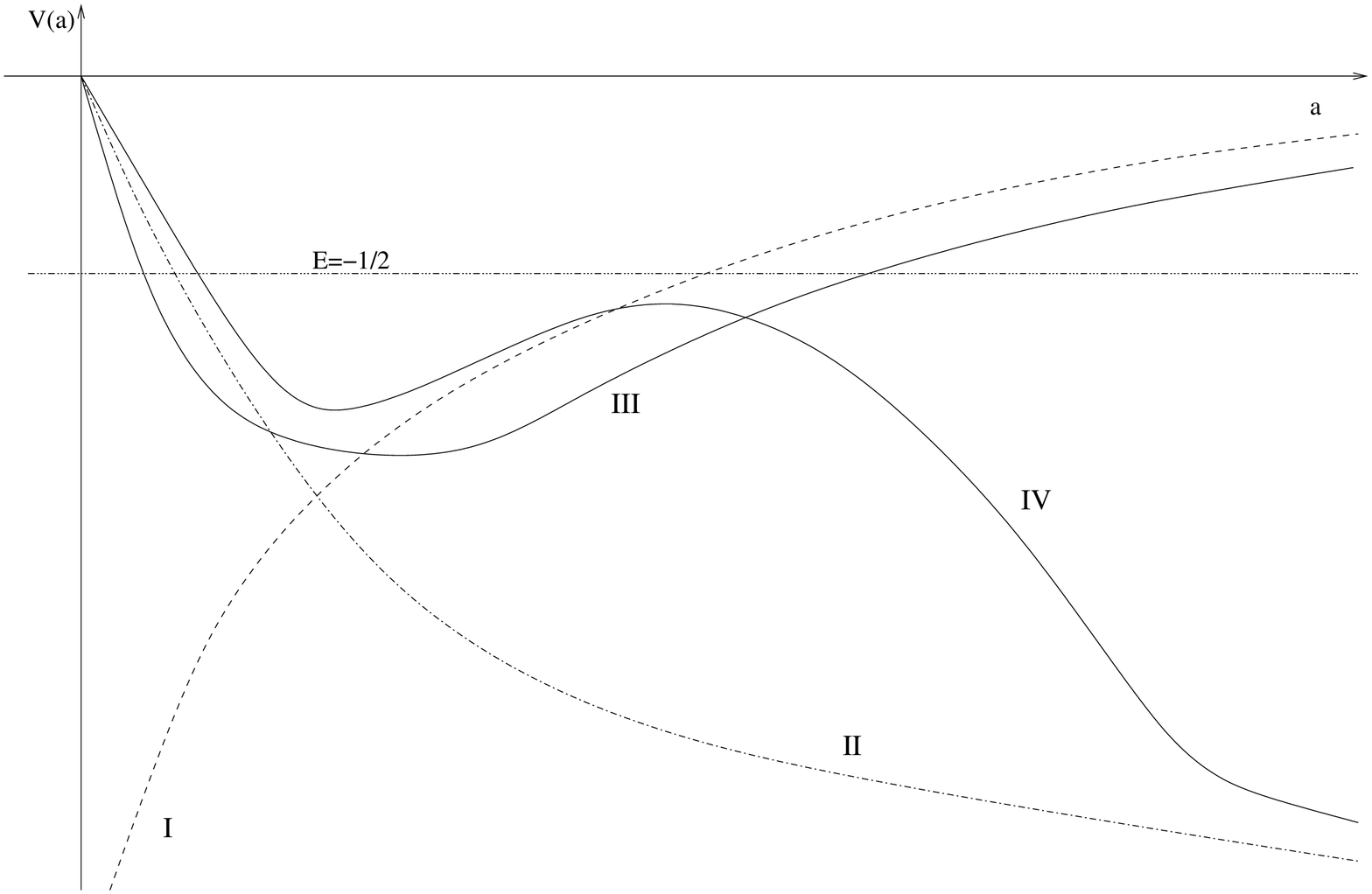,width=80mm}{Schematic depiction of
several hypothetical effective potentials $V(a)$ associated with
different cosmological evolutions.\label{violation}}

For matter with a constant equation of state the {\small SEC}
implies that the cosmological acceleration is negative. Curve~I on
figure~\ref{violation} represents a typical potential $V(a)$
associated with matter respecting the {\small SEC}. Using the
point particle analogy, the line labelled $E=-1/2$ (we consider
the $k=+1$ case) represents the conserved energy. In this case the
only $a(t)$--trajectories which are kinematically allowed are
those beginning their evolution for
$a<a_{c}$.\footnote{Generically we refer to $a_{c}$ as the
point(s) where the curve $V(a)$ intersects the line $E=-1/2$.}
However they always lead to a turning point at $a=a_{c}$ followed
by period of contraction leading to a singularity as predicted by
Theorem~I of section~\ref{singular}. Curve~II on
figure~\ref{violation} is associated with matter violating the
{\small SEC} ($-1 \leq w <-1+2/n$). Cosmologies corresponding to
initial conditions fixed at some $t_{0}$ when $a>a_{c}$ will lead
to eternal expansion with positive acceleration. A feature of the
corresponding sources is that they can support a non--singular
gravitational bounce. Beginning the evolution for large $a$ the
solutions can contract to a minimum scale $a_{c}$ and then
re--expand. We can consider other non--singular geometries
supported by an equation of state which does not violate the
{\small SEC} at all times. For example, curve~IV can roughly be
divided into three regions. For small $a$ there is a positive
acceleration region which is followed by a region where the
{\small SEC} is satisfied. The transition from small to
intermediate values of the scale factor corresponds to positive
acceleration following by deceleration (negative acceleration) not
unlike the transition between the inflationary and the
radiation--dominated phases in the standard model. For large $a$
the equation of state depicted on curve~IV leads to a speed--up
following the radiation$\rightarrow$matter--dominated era. This
could correspond to the current observed accelerating state of the
universe \cite{supernovae}. We note that this picture is not
inconsistent with a bounce for small values of the scale factor.
This bounce could in principle be associated with the dynamics of
the inflaton. Curve~III is perhaps less relevant. There are then
two critical points where the geometry can bounce. Given
appropriate initial conditions, the corresponding spacetimes could
be going through many cycles of collapse and re--expansion without
developing singularities.

The conclusion of the over--simplified analysis performed in this
section is nevertheless quite general. In order to obtain a
homogeneous, isotropic and non--singular bouncing spacetime with
positive spatial curvature there must be three regions of positive
acceleration, \ie, the past ($t\rightarrow -\infty$), the bouncing
region and the future ($t\rightarrow +\infty$). A recent example
of this can be found in ref.~\cite{sen10} where a
tachyon--cosmological constant system is analyzed. Other examples
of such systems are described in sections~\ref{sec3} and \ref{spacelikeb}.

\subsection{The infinite throat}
\label{sec2}

In this section we consider flat ($k=0$) solutions associated with
a potential of the form (\ref{thepot}). We consider why, although
the sources can violate the {\small SEC}, the singularity theorems
can be applied and used to explain the existence of a spacelike
curvature singularity. Then in section~\ref{sec3} we find
non--singular $k=+1$ solutions containing a gravitational bounce.

The ($p+1$)--dimensional action for the system under study is \beq
\int d^{p+1}x \sqrt{-g} \left[ R
-\frac{1}{2}g^{\mu\nu}\partial_{\mu}\phi
\partial_{\nu}\phi - V(\phi)\right]. \eeq The equations of motion
are found to be \beq R_{\mu\nu} =
\frac{1}{2}\partial_{\mu}\phi\partial_{\nu}\phi+\frac{1}{p-1}g_{\mu\nu}V(\phi),\eeq
\beq \frac{1}{\sqrt{-g}}\partial_{\mu}
\left(\sqrt{-g}\partial^{\mu}\phi\right) -\frac{\partial
V}{\partial \phi}=0, \eeq which must be supplemented with a first
order constraint (\eg, $T_{tt} = G_{tt}$ for cosmological
applications).

To facilitate the obtention of analytical solutions we consider a
time--dependent homogeneous metric ansatz with a non--trivial
lapse function, \beq \label{hypermetric} ds^2 = - e^{2A(t)} dt^2 +
e^{2B(t)} d\Sigma_{k,p}^{2}, \eeq where the $p$--dimensional
Euclidean metric $d\Sigma_{k,p}^{2}$ is \beq \label{transverse}
d\Sigma_{p,k}^{2} = \frac{d\xi^{2}}{1-k\xi^{2}} + \xi^{2}
d\Omega_{p-1}^{2}. \eeq The spatial metric $d\Omega_{p-1}^{2}$ is
that associated with a unit ($p-1$)--dimensional hypersphere. For
$k=+1$ expression (\ref{transverse}) is the unit metric on
$S^{p}$. For $k=-1$ the unit metric is the hyperbolic
$p$--dimensional space $H^{p}$. Using eq.~(\ref{hypermetric}) and
assuming the geometry is supported by a homogeneous scalar field,
$\phi=\phi(t)$, the relevant equations of motion become \beq
\ddot{B} + \dot{B}\left( p \dot{B} - \dot{A} \right) =
\frac{1}{p-1}e^{2A}V(\phi) - (p-1)k e^{2(A-B)}, \eeq \beq
\ddot{\phi} + \dot{\phi}\left( p\dot{B}-\dot{A} \right) +
e^{2A}\frac{\partial V}{\partial \phi}=0, \eeq with the Friedmann
constraint \beq \frac{p(p-1)}{2}\left( \dot{B}^{2} + k e^{2(A-B)}
\right) = \frac{1}{4}\dot{\phi}^{2} + \frac{1}{2} e^{2A} V(\phi).
\eeq

\EPSFIGURE[r]{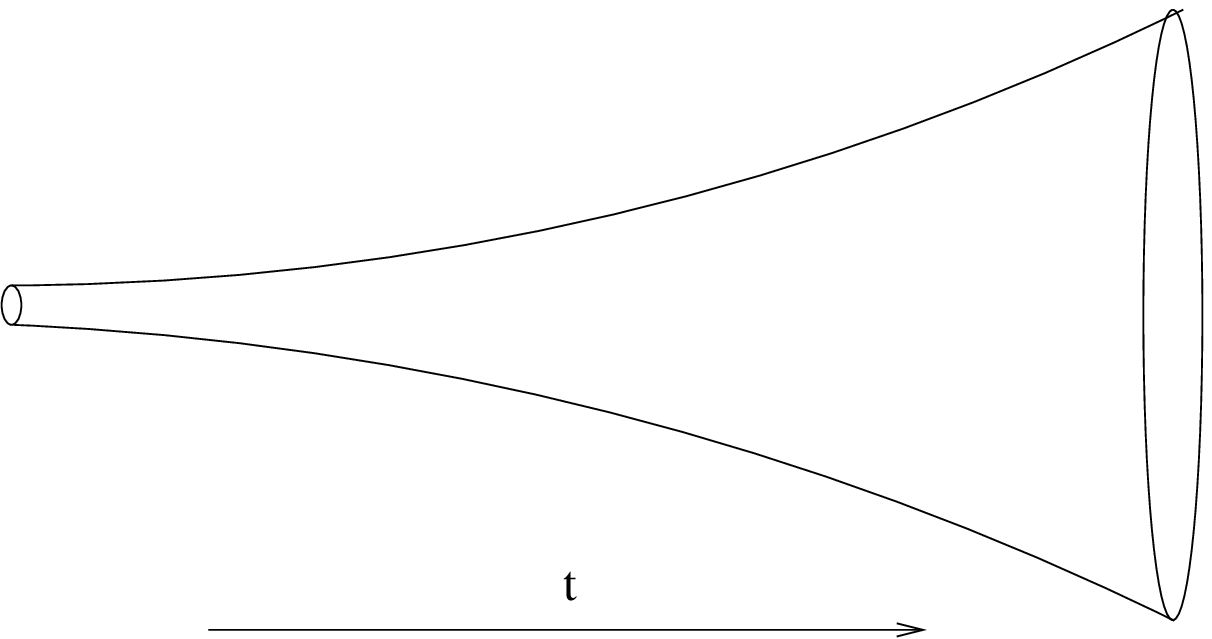,width=70mm}{Schematic depiction ofthe
infinite cosmological throat for $s=-1$. The volume function
$V(t)$ vanishes as $t\rightarrow -\infty$ which corresponds to a
singular spacelike horizon.\label{sing}} It is always possible to
use a change of variables of the form $t=t'(t)$ allowing us to
choose a convenient form for the lapse function. In what follows
we therefore make the gauge choice $A=pB$ which simplifies the
equation of motion for the scalar field. Let us then define the
volume function \beq V(t) = e^{2pB}, \eeq which is related to the
actual constant--time volume of the spatial sections through \beq
\int d^{p}x \sqrt{-g} = e^{2pB} \int \sqrt{\p g}, \eeq where $\p
g$ is the determinant associated with the Euclidean metric
(\ref{transverse}). The acceleration\footnote{It is crucial to
realize that this is not the acceleration measured using the
so--called cosmological time. This will be clarified later in this
subsection.} associated with the volume of the spacetime is \beq
\label{volumeacc} \ddot{V}(t) = 2p V(t) \left( \ddot{B} + 2p
\dot{B}^{2} \right). \eeq In the $A=pB$ gauge the dynamics of the
gravi--metric field $B(t)$ is governed by the equation \beq
\ddot{B} = -(p-1)k e^{2(p-1)B} + \frac{1}{p-1} e^{2pB} V(\phi).
\eeq It is clear that for $k=0$ and $V(\phi)>0$ the solutions have
a volume with eternal positive acceleration. This is also true
when considering hyperbolic foliations. Conversely for $k=+1$ the
curvature term contributes in such a way as to favor negative
acceleration. Analytic solutions can be obtained for $k=0$ (the
case $k=+1$ is treated in section~\ref{sec3}) and a potential of
the form (\ref{thepot}). The equations of motion are then \beq
\label{e1} \ddot{B} - \frac{\Lambda}{p-1} e^{2pB-\alpha\phi} = 0,
\eeq \beq \label{e2} \ddot{\phi} - \alpha \Lambda e^{2pB
-\alpha\phi} = 0, \eeq and the Friedmann constraint takes the form
\beq \label{e3} p(p-1)\dot{B}^{2} = \frac{1}{2}\dot{\phi}^{2} +
\Lambda e^{2pB-\alpha \phi}. \eeq It is then straightforward to
write down the solution for the scalar field in terms of the field
$B(t)$, \beq \phi(t) = \alpha (p-1) B(t) + c_{1}t + c_{2}, \eeq
where $c_{1}$ and $c_{2}$ are at this stage undetermined constant
parameters. Then we use the change of variables \beq h(t) =-\chi
B(t)-\alpha \left(c_{1}t + c_{2}\right) \eeq to write down
eq.~(\ref{e1}) in the form \beq \ddot{h} + \frac{\chi
\Lambda}{p-1} e^{h} = 0, \eeq where $\chi=\alpha^{2}(p-1) - 2p$.
For $\chi>0$, the solution to the latter equation is \beq h(t) =
\ln\left(\frac{p-1}{2\chi\Lambda c^{2}}\right) -
\ln\left(\cosh^{2}\left(\frac{t-t_{0}}{2c}\right) \right), \eeq
where $t_{0}$ and $c$ are constants of integration. For $\chi<0$
the solution becomes \beq h(t) = \ln\left(\frac{p-1}{2\chi\Lambda
c^{2}}\right) - \ln\left(\cos^{2}\left(\frac{t-t_{0}}{2c}\right)
\right), \eeq which develops curvature singularities in finite
time at $|t|=c\pi +t_{0}$. The reason for this is clear since for
$\chi<0$ the second derivative of $h(t)$ is necessarily positive
which implies that $\ddot{B}<0$. This is inconsistent with the
equations of motion.

\EPSFIGURE[r]{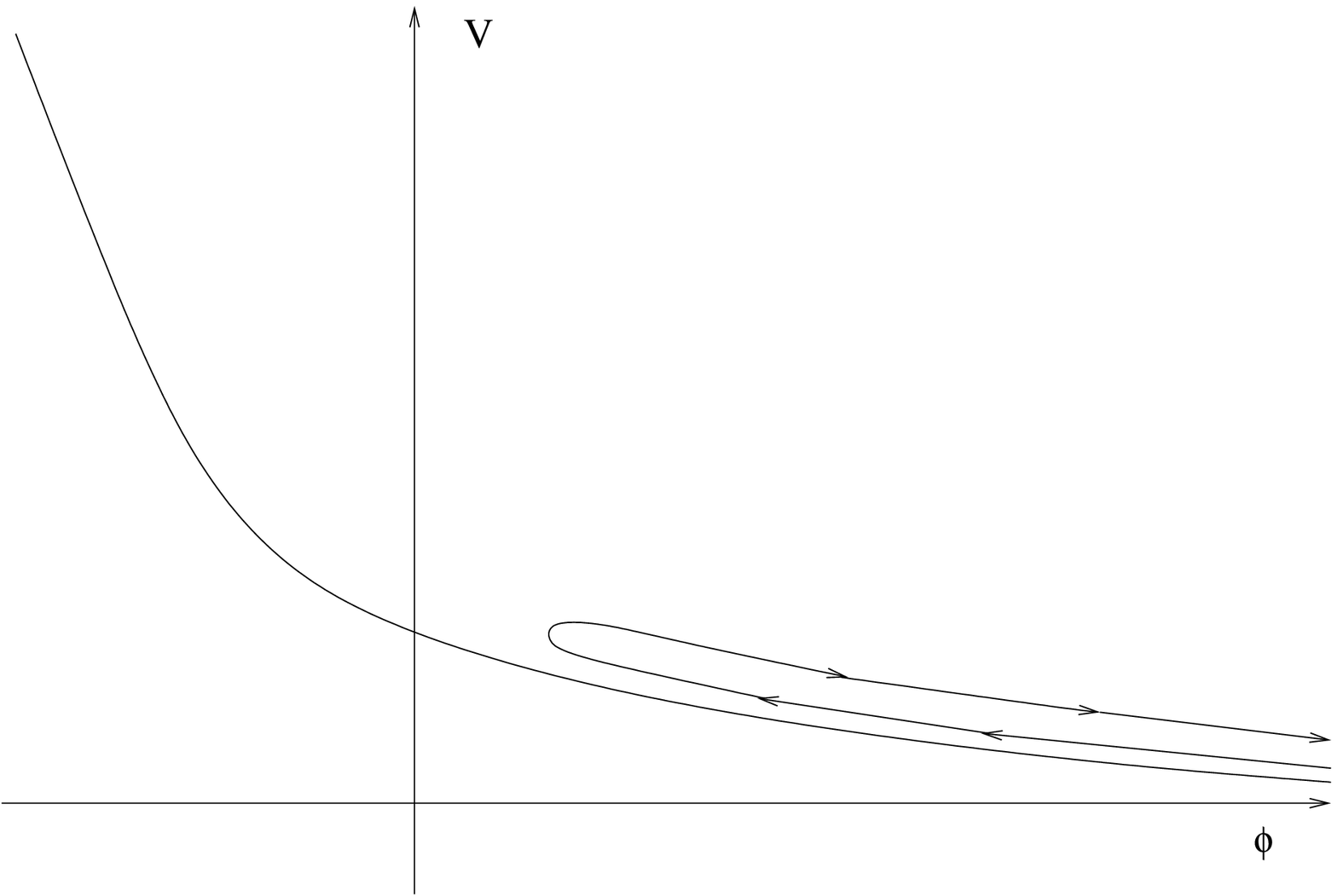,width=70mm}{Typical scalar field
trajectory associated with a bouncing cosmology. The potential is
of the form $V(\phi)=\Lambda e^{-\alpha\phi}$ where
$\alpha>0$.\label{evolfig}}
We consider further the solutions for which $\chi>0$, \ie, \beq
\alpha > +\sqrt{\frac{2p}{p-1}},\eeq where we chose the positive
root because of its relevance for string compactifications. The
features associated with $\alpha<0$ time--dependent backgrounds
are unchanged. To complete our analysis we need to make sure the
solutions found are consistent with the Friedmann constraint
(\ref{e3}). This leads to a relation between the integration
constants $c$ and $c_{1}$, \beq c^{2}c_{1}^{2} = \frac{p-1}{2p},
\eeq or \beq c_{1} = \frac{s}{c} \sqrt{\frac{p-1}{2p}},\eeq where
$s$ can either be $+1$ or $-1$. As will be shown below this sign
is important because it determines whether the spacetime is
expanding or contracting. Finally the solution takes the form \beq
B(t) = -\frac{1}{\chi} \left[ h(t) +s
\sqrt{1+\frac{\chi}{2p}}\left(\frac{t}{c}\right) +
\sqrt{\frac{\chi+2p}{p-1}}c_{2} \right], \eeq and the scalar field
is given by \beq \phi(t)= -\sqrt{\frac{2p(p-1)}{\chi}}h(t) +
s\sqrt{p-1}\left(
\frac{1}{\sqrt{2p}}-\sqrt{\frac{\chi+2p}{\chi}}\right)\left(\frac{t}{c}\right)
+ c_{2} \left(1-\sqrt{\frac{2p(\chi+2p)}{\chi}}\right). \eeq The
constant $t_{0}$ can be removed by the field redefinition
$t\rightarrow t-t_{0}$. There are therefore two physical
parameters characterizing the cosmological spacetimes, \ie, the
constants $c$ and $c_{2}$.

We now consider several important features associated with the
spacetime solutions found. First let us determine the general
behavior of the volume function $V(t)$. For $s=-1$ we find \beq
\label{thecurve} \lim_{t\rightarrow \pm\infty} \ln V(t) =
2\alpha_{\pm}\frac{t}{c}, \eeq where \beq \alpha_{\pm} =
\frac{1}{\chi}\left(\pm 1+ \sqrt{1+\frac{\chi}{2p}} \right) >
0.\eeq The volume function is therefore such that it evolves from
zero at $t=-\infty$ up to large values as $t\rightarrow +\infty$.
This is represented on figure~\ref{sing}. The fact that $V(t)$
vanishes as $t\rightarrow -\infty$ implies that this region
corresponds to an horizon. For $s=+1$ the spacetime is contracting
instead and the horizon is at $t=+\infty$. The relevant curvature
invariants are of the form \beq R =
\frac{2p}{V(t)}f_{1}(\dot{B},\ddot{B}), \eeq \beq
R_{\mu\nu}R^{\mu\nu} = \left(\frac{2p}{V(t)}\right)^{2}
f_{2}(\dot{B},\ddot{B}), \eeq \beq
R_{\mu\nu\rho\lambda}R^{\mu\nu\rho\lambda} =
\left(\frac{2p}{V(t)}\right)^{2} f_{3}(\dot{B},\ddot{B}), \eeq
with the square of the Weyl tensor obviously vanishing at all
times. The functions $f_{1}$, $f_{2}$ and $f_{3}$ (polynomials in
$\dot{B}$ and $\ddot{B}$) are finite for all values of the
parameter $t$. Therefore using eq.~(\ref{thecurve}) we find there
is a curvature singularity for $t\rightarrow -\infty$
(big--bang)\footnote{For $s=+1$ there is a big--crunch at
$t=+\infty$.} but that the curvature vanishes as $t\rightarrow
+\infty$.

The presence of singular points could not have been predicted
using singularity theorems because the {\small SEC} can be
violated in this case. For the solutions of interest the energy
condition ($R_{tt}\geq 0$) is equivalent to the inequality \beq
\label{inequality}
-\frac{1}{\cosh^{2}\left(\frac{t-t_{0}}{2c}\right)} +
\frac{2(p-1)}{\chi}\left(\tanh\left(\frac{t-t_{0}}{2c}\right)+
\sqrt{1+\frac{\chi}{2p}}\right)^{2} \geq 0. \eeq We are able to
show that this is always violated at least in some finite timelike
interval.\footnote{This leads to a period of positive cosmological
acceleration as was considered in refs.~\cite{townsend1,emparan}.}
In the asymptotic regions we have \beq
\lim_{t\rightarrow\pm\infty} R_{tt} = -\frac{2p}{\chi c^{2}}
e^{\mp \frac{t}{c}} + \left( \pm
1+\sqrt{1+\frac{\chi}{2p}}\right)^{2}. \eeq Since the first term
on the RHS asymptotically vanishes this shows that the {\small
SEC} is respected close to the horizon and also as $t\rightarrow
+\infty$. In fact the large $|t|$ behavior of the metric is found
to be \beq \label{larget} ds^{2} = -d\tau^{2} +
\left(\frac{p\alpha_{\pm}}{c}\right)^{\frac{2}{p}}
\tau^{\frac{2}{p}} d{\bf x}^{2}, \eeq where we have used the
change of variable \beq t = \frac{c}{p\alpha_{\pm}}\ln
\left(\frac{p\alpha_{\pm}}{c}\tau\right).\eeq The asymptotic
behavior (\ref{larget}) corresponds to the asymptotic equation of
state $w=1$. This explains why a singularity appears at
$t=-\infty$. Bounces are not allowed in this spacetime so in the
past the spacetime is expanding in a region where the {\small SEC}
is satisfied. Theorem~I therefore applies and predicts that
singular points will develop (in this case a genuine curvature
singularity at $\tau=0$ or $t=-\infty$). We have shown that
$R_{tt}$ extrapolates between two positive values between
$t=-\infty$ and $t=+\infty$. The question is whether there are
points in between where $R_{tt}<0$. The LHS of the inequality
(\ref{inequality}) is minimized for \beq \frac{t_{c}}{2c} = {\rm
arctanh}\, \left(-\frac{1}{\sqrt{1+\frac{\chi}{2p}}}\right). \eeq
Plugging this back in $R_{tt}$ we find that it is always negative
for $t=t_{c}$. This implies that the $k=0$ spacetimes found are
always associated with an intermediate period of positive
acceleration. The duration of this phase can be obtained by
studying further the inequality (\ref{inequality}).

\subsection{Bouncing spacetimes}
\label{sec3}

\EPSFIGURE[r]{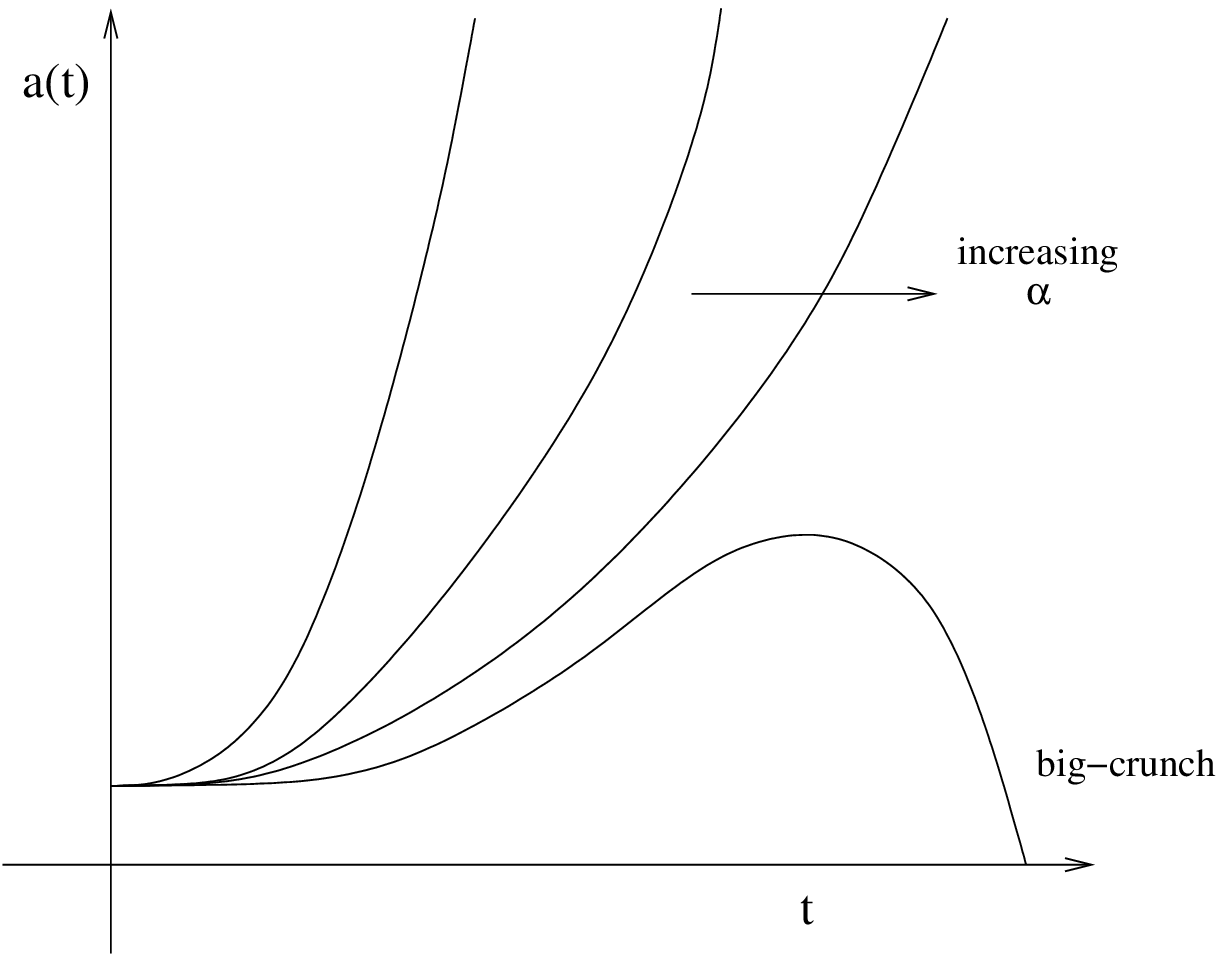,width=50mm}{Schematic depiction of the
scale factor behavior as a function the slope parameter $\alpha$.
\label{scalea1}} We have shown that the $k=0$ time--dependent
solutions with a potential of the form (\ref{thepot}) are always
singular. In this subsection we consider the corresponding $k=+1$
spacetimes. Our analysis is based on the FLRW metric ansatz \beq
ds^{2} = -dt^{2} +a(t)^{2} d\Omega_{p}^{2}. \eeq The equations of
motion are \beq \label{k1eom} \frac{\ddot{a}}{a} =
-\frac{1}{2p}\dot{\phi}^{2} + \frac{\Lambda}{p(p-1)}
e^{-\alpha\phi},\eeq \beq \ddot{\phi} =
-p\frac{\dot{a}}{a}\dot{\phi} + \alpha\Lambda e^{-\alpha\phi},
\eeq with the Friedmann constraint \beq p(p-1)\left[
\left(\frac{\dot{a}}{a}\right)^{2} + \frac{1}{a^{2}}\right] =
\frac{1}{2}\dot{\phi}^{2} +\Lambda e^{-\alpha\phi}. \eeq In
particular we investigate cosmological solutions containing a
gravitational bounce. Our results are obtained by solving the
system of differential equations numerically. The strategy is to
exploit the fact that both the equations of motion and the
boundary conditions are time--reversal symmetric. The latter are
chosen in order that the bounce occurs at $t=0$, \beq \label{bc3}
\dot{a}(0) = 0 = \dot{\phi}(0), \;\;\; \phi(0) = \phi_{0}. \eeq
The first order constraint imposes that the size of the $t=0$
Cauchy surface is fixed by \beq a(0) = \sqrt{\frac{p(p-1)}{\Lambda
e^{-\alpha\phi_{0}}}}. \eeq The type of scalar field trajectories
leading to a bouncing spacetime is depicted on
figure~\ref{evolfig}. The scalar rolls from $\phi=+\infty$,
reaches a maximum at $\phi=\phi_{0}$ and then rolls down again
toward small values of the potential. The bounce of the spatial
sections occurs precisely at the turning point for the scalar
field. We solved the equations numerically to find the geometry
corresponding to the roll down from $\phi=\phi_{0}$ at $t=0$ to
$\phi=+\infty$ at $t=+\infty$. The other half of the solution,
\ie, the past, is simply the time--reversed version of the $t>0$
solution. \EPSFIGURE[r]{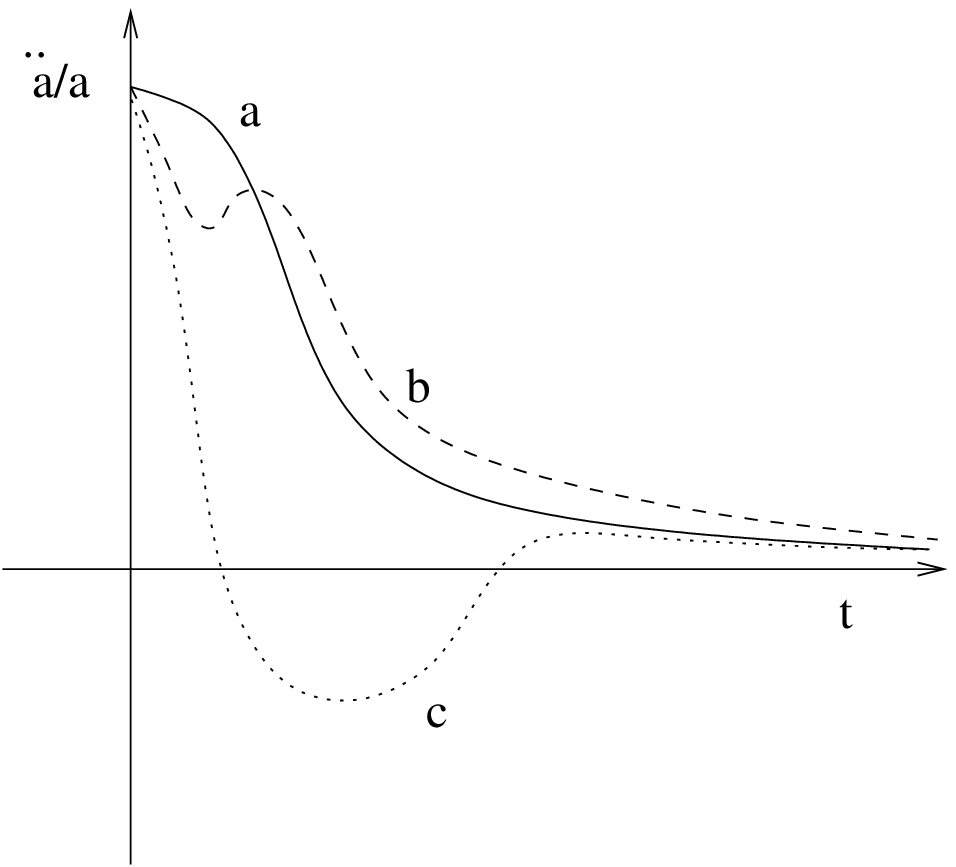,width=50mm}{Schematic
representation of the cosmological acceleration associated with
some $k=+1$ solutions supported by a scalar field in a positive
exponential potential.\label{acceleration}}

There are two important parameters in the system with boundary
conditions (\ref{bc3}). The combination $\Lambda
e^{-\alpha\phi_{0}}$ characterizes the height of the potential
when the scalar field is released from rest at $t=0$. However the most
relevant parameter (in terms of whether or not non--singular solutions exist)
is the dimensionless quantity
\beq \alpha = \left|\frac{\partial \ln V(\phi)}{\partial \phi}\right|,
\eeq which determines the slope of the potential.
Without loss of generality we fixed $\Lambda=1$ and
varied $\phi_{0}$ when scanning through all possible
solutions. We were interested in determining
what values of $\Lambda e^{-\alpha\phi_{0}}$ and $\alpha$ lead to
non--singular evolutions. Secondly, we wanted to study the nature
of the cosmological acceleration associated with the spatial sections.

It is relatively easy to generate solutions with a bounce. However
most of those develop singular points while the scalar field rolls
down the potential. We find there is a critical value
$\alpha=\alpha_{c}$ above which the bouncing spacetimes always
develop a curvature singularity. This feature is independent of
what values of $\phi_{0}$ and $\Lambda$ are chosen. This is
illustrated on figure~\ref{scalea1}. For $\alpha<\alpha_{c}$ we
obtain non--singular bouncing geometries associated with
asymptotic phases (large $|t|$) having positive acceleration.
Smaller values of $\alpha$ lead to larger values of asymptotic
acceleration as shown on figure~\ref{scalea1}. The critical value
for $p=3,4,5,6,7,8,9$ corresponds respectively to roughly
$10\alpha_{c}=6,5,4,3,3,2,2$.

Inspecting eq.~(\ref{k1eom}) we see that the contribution of the
scalar field to the acceleration is negative--definite while the
form field contributes a positive--definite term.\footnote{This
can of course be traced back to the fact that space--filling form
fields violate the {\small SEC}.} Which contribution ultimately
dominate determines the fate of the spacetime for large $|t|$. Let
us consider the $t>0$ case. Typically what happens for
$\alpha>\alpha_{c}$ is that the kinetic term becomes dominant and
drives the spacetime into a phase of contraction. Then since the
{\small SEC} holds Theorem~I from section~\ref{singular} is
applicable which supports our finding that singular points appear
in the future. This depressing state of affair does not persist
for $\alpha<\alpha_{c}$. In this case the contribution of the
potential to the acceleration dominates asymptotically and the
spacetime never enters a phase of contraction. This implies that
the {\small SEC} is violated for large $|t|$, \ie, the scale
factor will behave like \beq \lim_{t\rightarrow\pm\infty} a(t) =
t^{m}, \eeq where $m>1$. Among other things this implies that
while $\ddot{a}/a$ and $\dot{a}/a$ vanish asymptotically,
quantities such as $\dot{a}$ are unbounded.\footnote{This is of
course not worrisome since $\dot{a}$ is not an observable.} In
fact the effective potential $V(a)$ (see section~\ref{sec1}) will
be of the form of either curve~II or curve~IV on
figure~\ref{violation}.

We have determined that the non--singular $k=+1$ solutions have
three phases of positive acceleration: the past, the bounce and
the future. The qualitative behavior for the intermediary phases
can take three different forms. Curve~a) on
figure~\ref{acceleration} represents a common signature where the
acceleration remains positive during the whole evolution. In some
instances $\ddot{a}/a$ always remains positive but develops a kink
in finite time as represented by curve~b). Interestingly, for some
values of the parameters this kink drops below zero which
corresponds to the spacetime entering a regime of negative
acceleration. For example this happens for the case $p=4$,
$\alpha=2/5$, $\Lambda=1$ and $\phi_{0}=10$. The corresponding
acceleration is depicted by curve~c).\footnote{The form of the
associated effective potential would be that represented by
curve~IV on figure~\ref{violation}.} This behavior is interesting
since it corresponds to a phase of large positive acceleration
(close to $t=0$) followed by a phase of negative acceleration with
a future characterized by a small (compared to that at $t=0$)
positive acceleration. The large $|t|$ contribution of the spatial
curvature is negligible with respect to the energy density
($a^{-2}/\rho \simeq 0$). This is reminiscent of our own universe
which begins with an inflationary phase followed by radiation--
then matter--dominated era. The observed late time behavior is
that of an accelerating spacetime with equation of state close to
$w=-1$ \cite{supernovae}.
\EPSFIGURE[r]{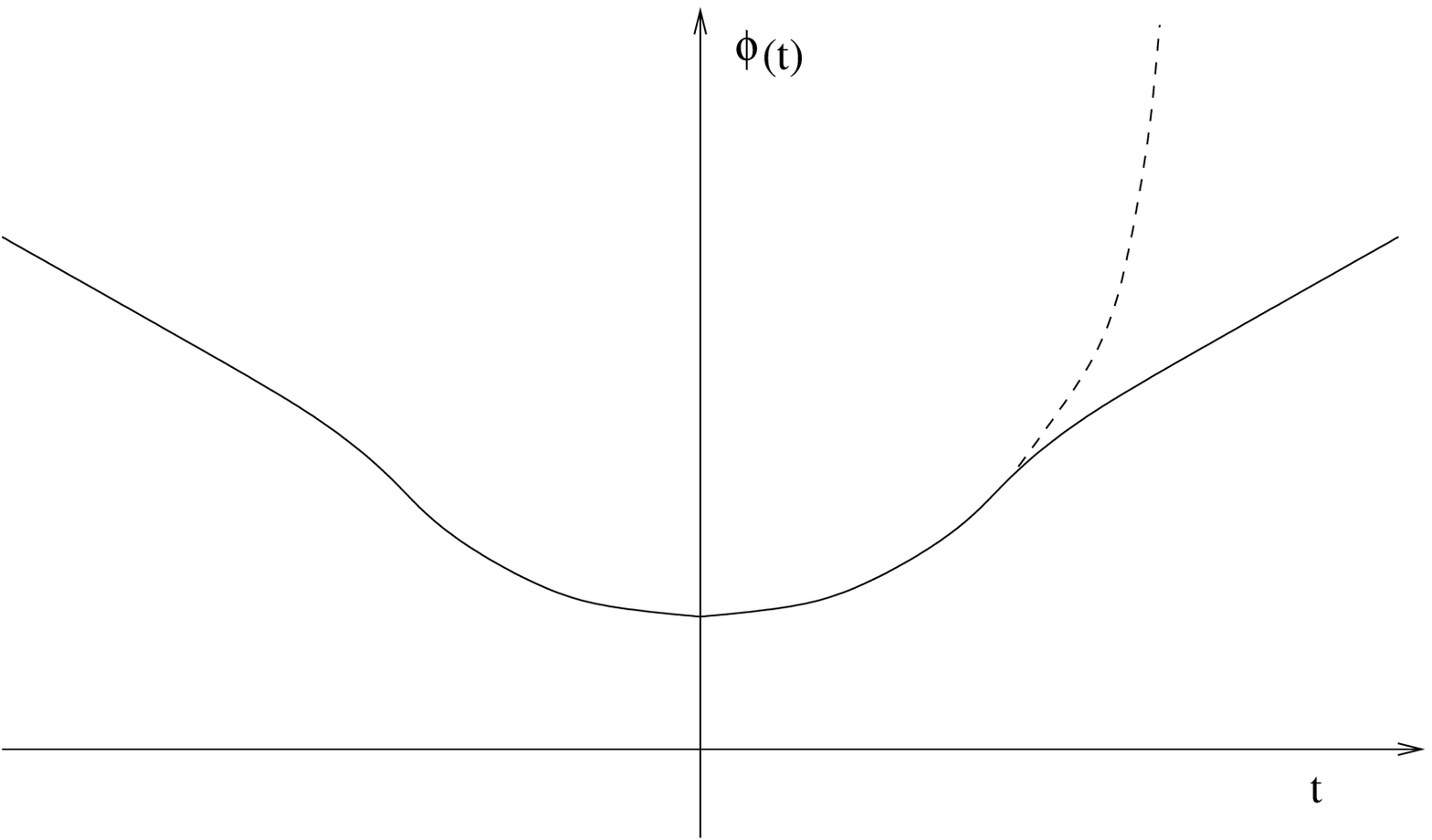,width=50mm}{Typical evolution of a scalar
field supporting a bouncing spacetime with $k=+1$. The dashed
curve represents an evolution leading to a curvature
singularity.\label{curvescalar}}

\section{Supergravity applications}
\label{applications}

In the previous section we found time--dependent gravitational
solutions with periods of positive acceleration. The $k=0$
solutions are singular while some of the $k=+1$ solutions are
regular because of the presence of a bounce. In this section we
consider whether or not these geometries can be embedded in
ten--dimensional (more precisely Type~IIA and Type~IIB) and
eleven--dimensional supergravity theories. However the resulting
higher--dimensional geometries will always be singular. This is
expected because all time--dependent solutions of ten-- and
eleven--dimensional supergravity contain singular points since the
gravitational sources do not violate the {\small SEC}.\footnote{As
pointed out earlier there are a few exceptions. In
section~\ref{spacelikeb} we consider systems associated to those.}

Suppose the regular ($p+1$)--dimensional bouncing solutions found
in section~\ref{sugra} can be embedded in a supergravity theory.
Then the non--singular character in ($p+1$) dimensions is only an
illusion of the compactification scheme. In fact based on
Theorem~I in section~\ref{singular} the uplifted geometry must
contain singular points associated with, for example, a breathing
mode driving the higher--dimensional spacetime toward
gravitational collapse (small spacetime volume) in a region where
the lower--dimensional scale factor increases. In other words, in
the past and/or the future of a lower--dimensional bounce, the
volume of the full spacetime will be driven toward gravitational
collapse although the four--dimensional geometry appears to be
non--singular.

The idea that such exotic effects as positive acceleration and
gravitational bounces are geometrical effects in gravitational
theories with more than four dimensions can have far--reaching
consequences for conventional cosmology. For example this
geometrical effect could change our perspective on issues related
to the hypothetical initial singular state sometimes associated
with the big--bang. An interesting hypothesis is that rather than
expanding out of a singular state our universe simply bounced off
after having reached a finite size and expanded toward its current
state. As shown in section~\ref{bounces} this is only physically
realizable in theories with sources violating the {\small SEC}.
The geometrical effect described above provides us with enough
leeway to conjecture that bounces are not excluded in
higher--dimensional models respecting the {\small SEC}. However
since the singularity theorems predict the apparition of singular
points the regular nature of the lower--dimensional spacetime is
an illusion. This is still useful however because in principle the
singular region can be made to appear arbitrarily far in the past
of the bouncing region. For a lower--dimensional observer this
would appear as though the big--bang singularity has been resolved
by a bounce.

\subsection{Flux compactifications}

In this subsection we consider $d=(m+1)=(p+1+n)$--dimensional
spacetimes of the form \beq \label{bigmetric} ds^{2} =
G_{IJ}dx^{I}dx^{J} = e^{\xi\psi} g_{\mu\nu}dx^{\mu}dx^{\nu} +
e^{2\psi} d\Sigma_{n,\sigma}^{2}, \eeq where $g_{\mu\nu}$ is the
metric associated with a ($p+1$)--dimensional Lorentzian spacetime
and $I,J=0,\,...\, ,m$. The scalar field $\psi$ can be regarded as
a breathing mode for the maximally symmetric Euclidean manifold
with curvature $\sigma=-1,0$ or $+1$. The study of models with a
transverse space associated with richer symmetry groups is beyond
the scope of our work (see ref.~\cite{ulf}). The theories of
interest to us are the ten-- and eleven--dimensional
supergravities so we consider a general Einstein frame action of
the form \beq \label{beginning} S = \frac{1}{16\pi G_{(m+1)}} \int
d^{m+1}x \sqrt{-G} \left[ {}^{\scriptscriptstyle (m+1)} R -
\frac{1}{2}\partial_{I}\phi\partial^{I}\phi -
\frac{1}{2(p+1)!}e^{a\phi} F_{[p+1]}^{2} \right], \eeq where
$G_{(m+1)}$ is the $d$--dimensional Newton constant, $\phi$ is the
dilaton field (absent for $m=10$), and $F_{[p+1]}$ is the field
strength associated with the Ramond--Ramond (RR) form fields
$C^{p}$. In ten--dimensional supergravity the dilaton coupling is
$a=(4-p)/2$\footnote{This ensures that the dilaton does not couple
to the RR fields kinetic term in the string frame.} and our
notation is
$F_{[p+1]}^{2}=F_{\mu...\mu_{p}}F^{\mu...\mu_{p}}$.\footnote{We
consider only form fields having their indices along $p$ spatial
directions on the Lorentzian manifold.}

We now write down, starting with expression (\ref{beginning}), an
effective action for gravity on the ($p+1$)--dimensional
Lorentzian sub--manifold with metric $g_{\mu\nu}$. Before
proceeding with the dimensional reduction we solve the equation of
motion associated with the form field. For the ansatz
$F_{\mu\mu_{1}...\mu_{p}}=\epsilon_{\mu\mu_{1}...\mu_{p}}A(x^{\nu})$
the equation of motion \beq \partial_{I}\left(
\sqrt{-G}e^{a\phi}F^{I I_{1}...I_{p}}\right) =0 , \eeq is solved
for \beq A(x^{\nu}) = C e^{-n\psi-a\phi}, \eeq where $C$ is a
constant. Using this result and the conventional Kaluza--Klein
ansatz (see ref.~\cite{carroll} for a modern treatment) we find
the dimensionally reduced action \beq \label{actionmain} S =
\frac{1}{16\pi G_{\scriptscriptstyle N}} \int d^{p+1}x \sqrt{-g} \left[ \inp
R - \frac{n(n+p-1)}{p-1}\partial^{\mu}\psi\partial_{\mu}\psi -
\frac{1}{2}
\partial^{\mu}\phi\partial_{\mu}\phi - V(\psi,\phi) \right], \eeq where
the effective potential for the dilaton and the breathing mode is
given by \beq V(\psi,\phi) = \frac{C^{2}}{2} e^{-\left(a\phi +
\frac{2np}{p-1}\psi\right)} - \n R
e^{-2\left(1+\frac{n}{p-1}\right)\psi}. \eeq In order to obtain a
conventionally normalized Ricci term we used $\xi = -2n/(p-1)$ and
the relation \beq \frac{1}{16\pi G_{\scriptscriptstyle N}} =
\frac{1}{16\pi G_{(m+1)}}\int d^{n}y \sqrt{\n g}, \eeq relating
the ($p+1$)--dimensional Newton constant ($G_{\scriptscriptstyle
N}$) to the higher--dimensional gravitational scale. The kinetic
term for the breathing mode in eq.~(\ref{actionmain}) is not
canonically normalized. This is fixed by applying \beq \psi
\rightarrow \sqrt{\frac{p-1}{2n(n+p-1)}}\bar{\psi} \eeq which
changes the form of the potential to \beq \label{finalpotential}
V(\bar{\psi},\phi) = \frac{C^{2}}{2} e^{-\left(a\phi
+\frac{\sqrt{2n}p}{\sqrt{(p-1)(n+p-1)}}\bar{\psi}\right)} - \n R
e^{-2\left(1+\frac{n}{p-1}\right)\sqrt{\frac{p-1}{2n(n+p-1)}}\bar{\psi}}.
\eeq

\subsection{Effective potential and scalar field dynamics}

The form of the exponential potential term for the dilaton in
eq.~(\ref{finalpotential}) is determined by the magnitude and sign
of $a$. If $a$ is negative ($p>4$ in Type~IIA and Type~IIB
supergravity) then the solutions will evolve between two
asymptotic regions where the dilaton contribution to the potential
asymptotically vanishes, \ie, as $\phi\rightarrow -\infty$ when
$t\rightarrow\pm\infty$. This implies that the geometry
extrapolates between two regions where the string coupling,
$g_{s}=e^{2\phi}$, is small. If $a$ is positive ($p<4$ in
supergravity) the corresponding solutions will include asymptotic
regions where the string coupling is unbounded.

The spatial volume of the Cauchy surfaces associated with the
metric (\ref{bigmetric}) is controlled by the quantity \beq
\label{thevolume} \int d^{m}x \sqrt{-G} = \int d^{p}x d^{n}y
\sqrt{\n g} \left[ \sqrt{-g} e^{-\frac{2n}{p-1}\bar{\psi}}
\right], \eeq where the time--dependent contributions are inside
the square--brackets. Inspection of eq.~(\ref{finalpotential})
shows that the argument of the exponential functions in
$\bar{\psi}$ is negative (for positive $\bar{\psi}$). Typically
this implies that all (potentially non--singular) time--dependent
solution are such that \beq \lim_{t\rightarrow\pm\infty} \psi(t) =
+\infty. \eeq This condition is physical since, for example, if it
is not satisfied the potential energy of the scalar could become
unbounded. This means, as suggested by eq.~(\ref{thevolume}), that
the contribution of the breathing mode to the evolution of the
Cauchy surfaces favors a contraction to small size both in the
asymptotic past and future.

Of course this last comment does not take into consideration the
dynamics associated with the effective ($p+1$)--dimensional
cosmology (\ie, the $\sqrt{-g}$ contribution to
expression~(\ref{thevolume})). The time--dependence of
$g_{\mu\nu}$ can, in principle, be non--singular and, for example,
include a gravitational bounce. Of course this type of behavior is
in principle allowed because the {\small SEC} is effectively
violated in the lower--dimensional theory. However the
higher--dimensional theory satisfies the latter condition which
implies the presence of singular points. This strongly suggest
that the breathing mode will always drive the full spacetime
toward catastrophic gravitational collapse either (or both) in the
past and the future.

\subsection{The arduous ascension}

We study whether or not the solutions found in section~\ref{sugra}
can be embedded in Type~IIA and Type~IIB supergravity. These solutions
correspond to truncations of the theories with the action
(\ref{beginning}) such that a single exponential term
survives in the effective potential eq.~(\ref{finalpotential}). An
example consists in considering the transverse space to be a
$n$--dimensional torus ($\n R=0$). We also assume for simplicity
that the dilaton is turned off and that $a=0$. A generalization
including the dilaton can be found in appendix~\ref{scalarstuff}.

We begin by embedding the $k=0$ solutions found in
section~\ref{sec2}. The solutions that can be consistently
uplifted are those for which \beq \alpha =
\frac{\sqrt{2n}p}{\sqrt{(p-1)(n+p-1)}}, \;\;\;\; \Lambda =
\frac{C^{2}}{2}. \eeq We consider only the expanding infinite
throat ($s=-1$) which is associated, from the lower--dimensional
perspective, to a singular horizon at $t=-\infty$. The important
result is that the asymptotic volume of the uplifted solutions
behaves like \beq \lim_{t\rightarrow\pm\infty} \ln \left[
\sqrt{-g}e^{-\frac{2n}{p-1}\bar{\psi}} \right] = \kappa_{\pm}
\frac{t}{c} = \left( 2\alpha_{\pm}(p-\alpha n) + n
\sqrt{\frac{2}{p(p-1)}} \right) \frac{t}{c}. \eeq The sign of the
constants $\kappa_{+}$ and $\kappa_{-}$ is the determining factor
here. We find, for $d=10$ and $d=11$, that $\kappa_{-}$ is
positive for all relevant values of $p$. This implies that the
point $t=-\infty$ is also singular in the uplifted geometry. Now
for $d=10$ and for $d=11$ we find that $\kappa_{+}$ becomes
negative respectively for $p\geq 6$ and $p\geq 7$. The uplifted
geometries for which $\kappa_{+}>0$ are expanding and
non--singular in the future while those with $k_{+}<0$ enter a
phase of contraction in the future. This implies that singular
points will appear as predicted by the cosmological singularity
theorems. We have also verified explicitly that curvature
singularities appear in the future of the $\kappa_{+}>0$
solutions. The resulting uplifted geometries are therefore
associated with both a big--bang and a big--crunch singularity.
This is an example where the non--singular nature of a
lower--dimensional spacetime (the future of the effective
($p+1$)--dimensional spacetime) is only an illusion of the
compactification scheme considered.

The compactifications considered in this section do not lead to
effective theories allowing us to embed the non--singular bouncing
solutions found in section~\ref{sec3}. It is not excluded however
that those could be embedded for compactifications on manifolds
associated with more interesting symmetries. The singular
$(p+1)$--dimensional bouncing solutions can easily be embedded in
ten-- or eleven--dimensional supergravity. It is conceivable that
one of the two singularities (either the big--bang or the
big--crunch) is actually lifted and disappears from the
higher--dimensional point of view.

It would be very interesting to study further the
($p+1$)--dimensional gravitational system associated with a
potential (\ref{finalpotential}) for which $\n R>0$ and $C\neq 0$.
In this case the curvature term will favor negative acceleration.
For the $k=+1$ solutions this is a dangerous contribution if it
comes to dominate over the {\small SEC} violating contributions
during a phase of expansion. The dynamics of this system is the
result of a constant competition between the curvature and the
flux terms. If non--singular bouncing solutions are found the
$t>0$ region could correspond to a phase of negative acceleration
between two phases of positive acceleration. It would be
interesting to see if the phase of positive acceleration around
$t=0$ can be used as a realistic model of inflation. Then it would
be interesting to verify whether the asymptotic region of positive
acceleration can resemble closely enough the phase of positive
acceleration that is currently observed in our universe
\cite{supernovae}.

\section{Spacelike branes}
\label{spacelikeb}

We conclude this paper by considering an enigmatic class of
time--dependent solutions in string theory: spacelike branes. The
s--branes were conjectured to be the phenomenon associated with
the creation of a D--brane from a closed string vacuum and its
subsequent decay into closed strings \cite{strominger1}. These
objects were studied from different perspectives. From a classical
gravitational point of view, s--branes should correspond to
non--singular time--dependent backgrounds extrapolating (in time)
between two asymptotically FLRW spacetimes. Our main results here
are twofold. First we close a gap left opened in the literature
\cite{walcher,peet2} by showing that the space--filling s8--brane
is singular. Then, perhaps more importantly, we propose a natural
mechanism to resolve the singularities associated with the
supergravity s--branes \cite{gutperle,myerssbrane,peet,walcher}.
The resulting non--singular s--brane configurations would be
asymptotically dS.

The conventional approach to studying s$(p-1)$--branes in the
context of type~IIA,B supergravity \cite{strominger1,gutperle,
myerssbrane} consists in considering a geometry of the form \beq
\label{metricsbrane} ds^{2} = -dt^{2} +
a(t)^{2}d\Sigma_{p,k_{1}}^{2} + R(t)^{2} d\Sigma_{9-p,k_{2}}^{2},
\eeq coupled to homogeneous time--dependent dilaton and RR form
field $C^{p}$. The symmetry group for the s--brane was argued in
ref.~\cite{strominger1} to be that associated with $k_{1}=0$ and
$k_{2}=-1$ but we do not adopt this convention in the following
analysis and consider, for instance, spherical branes as well. The
Euclidean sub--manifold with scale factor $a(t)$ is the
worldvolume of the unstable brane. As pointed out earlier any
excitation of the sources in this problem, \ie , the dilaton and
the RR field, is always such that the {\small SEC} is satisfied.
This implies that the supergravity solutions for s--branes are
always singular \cite{walcher} since the cosmological theorems
reviewed in section~\ref{singular} are applicable. However, as
pointed out in refs.~\cite{walcher,peet2} there is one exception:
the s8--brane.

Recently so--called non--singular s--brane solutions were
presented in the literature by considering some analytic
continuations of known static black hole solutions \cite{alex}
(see also refs.~\cite{wang,cliffb}). In upcoming work
\cite{robbins} we consider in what sense these solutions evade the
higher--dimensional singularity theorems presented in
section~\ref{singular}.

\subsection{The s8--brane and the rolling tachyon}

The space--filling s8--brane is different from the other spacelike
branes if only for the fact that it has no transverse spatial
directions. If space--filling unstable branes are physical objects
then the coupling of the associated open string tachyon to the RR
field is of the form \beq S_{{\scriptscriptstyle WZ}} = \int f(T)
dT \wedge C^{9}, \eeq where the form of $f(T)$ can be found in,
\eg, ref.~\cite{kazumi}. The non--propagating form field $C^{9}$
is for example present in massive Type~IIA supergravity (see
ref.~\cite{polchinski}).

As pointed out earlier in this work spacetime filling form fields
(associated with a field strength with as many indices as there
are spacetime dimensions) violate the {\small SEC}. This is
interesting because it suggests that the s8--brane might be
non--singular since the cosmological singularity theorems are not
applicable in this case. We consider both the flat ($k=0$) and the
spherical ($k=1$) space--filling branes. For $k=1$ the solutions
we consider should correspond to a ten--dimensional bouncing
spacetime extrapolating between two asymptotically FLRW regions.
The $k=0$ solutions cannot go through a bouncing phase because,
according to the constraint (\ref{thebig}), $\dot{a}=0$ is
inconsistent with positive energy density. Non--singular solutions
could then take the form of a forever expanding or contracting
spacetime. Clearly this type of behavior is only realizable if the
dominating matter component in the system violates the {\small
SEC} in the asymptotic region where the volume of the spacetime
becomes small. This can almost immediately be excluded however
since, as shown in section~\ref{sec1}, as $a\rightarrow 0$ the
components violating the {\small SEC} become completely negligible
with respect to the other sources. These non--bouncing solutions
would perhaps be more relevant for the gravitational physics of
unstable branes henceforth referred to as half--s--branes.

The supergravity action associated with the s8--brane is the
expression (\ref{beginning}) with $m=9=p$ and $a=-5/2$. The
corresponding equations of motion are equivalent to that of a
single scalar field (the dilaton since there are not transverse
dimensions) coupled to gravity with a positive potential of the
form (\ref{thepot}) where $\alpha=-5/2$. $\Lambda$ is then a
parameter corresponding to the magnitude of the 9--form flux. In
sections~\ref{sugra} and \ref{applications} we have considered
explicitly only solutions with positive $\alpha$. The solutions
are unaltered for negative $\alpha$ but the distinction is
important in string theory. In fact because $\alpha$ is negative
for the s--branes of interest, the geometries would extrapolate
between vacua ($\phi\rightarrow\pm \infty$) where the string
coupling is very small ($g_{s}=e^{2\phi}$).

The s8--brane solutions were already studied in
section~\ref{sugra}. We found that the $k=0$ solutions are always
singular either in the past (big--bang solution) or in the future
(big--crunch solution). For the spherical s8--brane, bounces
actually occur but since $|a|=5/2>\alpha_{c}$ the corresponding
solutions are singular both in the past and the future of the
bounce. In other words it is the matter components respecting the
{\small SEC} that eventually comes to dominate before and after
the bouncing region. The apparition of singular points is then
predicted by the singularity theorems because the latter sources
always succeed in driving the spacetimes into a phase of
contraction (expansion) sometime in the future (past).

Refs.~\cite{buchel1,peet} suggested that the singularities found
above could be resolved by introducing in the system the most
relevant open string degree of freedom namely the tachyon. The
latter couples to gravity and the dilaton with a term of the form
(see \cite{peet} and references therein) \beq \label{braneaction}
S_{{\it brane}} = -T_{p} \int d^{p+1}x \, e^{-\phi} V(T)
\sqrt{-|{\cal P}G_{ab}+\partial_{a}T\partial_{b}T|} \, \delta({\bf
y}), \eeq where $V(T)=1/\cosh(T/\sqrt{2})$, ${\cal P}$ is the
pullback and the delta function localizes the unstable object in
the transverse spatial directions labelled ${\bf y}$. The form of
the potential and the regime of validity associated with the
action (\ref{braneaction}) are considered in ref.~\cite{kutasov}.
This minimum extension consisting in coupling the tachyon to the
massless closed string modes\footnote{This approach neglects the
effect of inhomogeneities which are very much relevant (see
ref.~\cite{kofman}).} was doomed from the start. In fact it is
shown in ref.~\cite{walcher} that the source (\ref{braneaction})
respects the {\small SEC} for $p\leq 6$. Based on the analysis
presented in this paper it is clear that such a contribution
cannot change the singular outcome associated with the
supergravity s--branes found in refs.~\cite{gutperle,myerssbrane}.

However for $p=6$ and $p=7$ the tachyon typically favors a short
period of positive acceleration whenever its time--derivative is
close to zero. Typically the tachyon extrapolates from a region
where $\dot{T}\approx 0$ (when the tachyon is close or at the top
of the potential $V(T)$) toward a late time asymptotic state
(tachyon matter) where $\dot{T}\rightarrow 1$ \cite{peet}. The
phase of positive acceleration can therefore occur only for a
short time after the tachyon starts rolling down. This can
certainly act as a gravitational source for a bounce. However for
the spherical s$8$--brane this cannot be used to resolve the
big--crunch and big--bang singularities we found earlier because
the period of positive acceleration is too short. Even worst, the
tachyon contribution at late and early time favors gravitational
collapse. Similar pessimistic comments apply to the s7--brane.
However a case deserving further study is that of the $k=0$ s7--
and s8--branes coupled to the tachyon. It is then conceivable that
the $\dot{T}\approx 0$ region could be associated with the
singular $t\rightarrow -\infty$ (for the big--bang solution)
region. This may actually insure a safe landing (as $t\rightarrow
-\infty$) for the scale factor since the cosmological singularity
theorems would not apply there.

\subsection{Unstable branes within branes}

As our final result in this paper we describe a general mechanism
by which the cosmological singularities associated with s--branes
could be resolved. This applies both to the supergravity s--branes
\cite{strominger1,gutperle, myerssbrane} and to s--brane gravity
fields supported by a rolling tachyon \cite{buchel1,peet}. Before
providing examples let us be more precise with respect to the
nature of our claims by stating a conjecture:

\begin{quotation} \noindent Non--singular
time--dependent spacetime solutions associated with supergravity
s--branes can only exist if the evolution of the fields takes
place in the presence of either stable D9--branes (D8--branes) in
Type~IIB (Type~IIA) supergravity, or, a more elaborate
configuration of lower--dimensional (perhaps smeared along some
directions) branes such that the corresponding gravitational
contribution violates the strong energy condition.
\end{quotation}

\noindent In other words we propose that, if the s--brane
supergravity backgrounds are somehow dual to a tachyonic open
string theory \cite{openclosed}, only in cases for which the end
point of the decay (and the point where brane creation begins) has
non--zero vacuum energy can the gravitational backgrounds be
non--singular. In fact we have seen in section~\ref{energy} that
tensile objects with spatial co--dimension zero or one are
gravitational sources violating the {\small SEC}. Adding an ingredient
like that to the usual supergravity system associated to
s--branes \cite{gutperle,myerssbrane} at least invalidates our
intuition based on the singularity theorems presented in
section~\ref{singular}. In the remaining of this section we
provide evidence that the conjecture actually holds.

\subsubsection{Asymptotically dS non--singular s--branes}

We begin by considering two very simple examples: (1) the Type~IIB
supergravity s7--brane in the presence of $N$ stable D9--branes
and, (2) the type~IIA s8--brane in the presence of D8--branes.
Evidence for the non--singular nature of the perhaps more
interesting lower--dimensional branes is provided in the next
section.

The action associated with the massless fields sourced by an
unstable 7--brane is expression (\ref{beginning}) with $m=9$,
$p=8$ and $a=-2$. The contribution from the D9--branes is in the
form of a ten--dimensional cosmological constant, \beq
\label{cosmoterm} S_{D9} = -\frac{1}{16\pi G_{10}} \int d^{10}x
\sqrt{-G} \Lambda_{b}, \eeq with \beq \Lambda_{b} = \frac{N
g_{s}}{(2\pi l_{s})^{3}}. \eeq The non--vanishing energy density
is simply $N$ times the D9--brane tension \cite{polchinski}. This
new ingredient is precisely what is needed in order to resolve the
IR (large $|t|$) singularities usually associated with s--branes.
\EPSFIGURE[r]{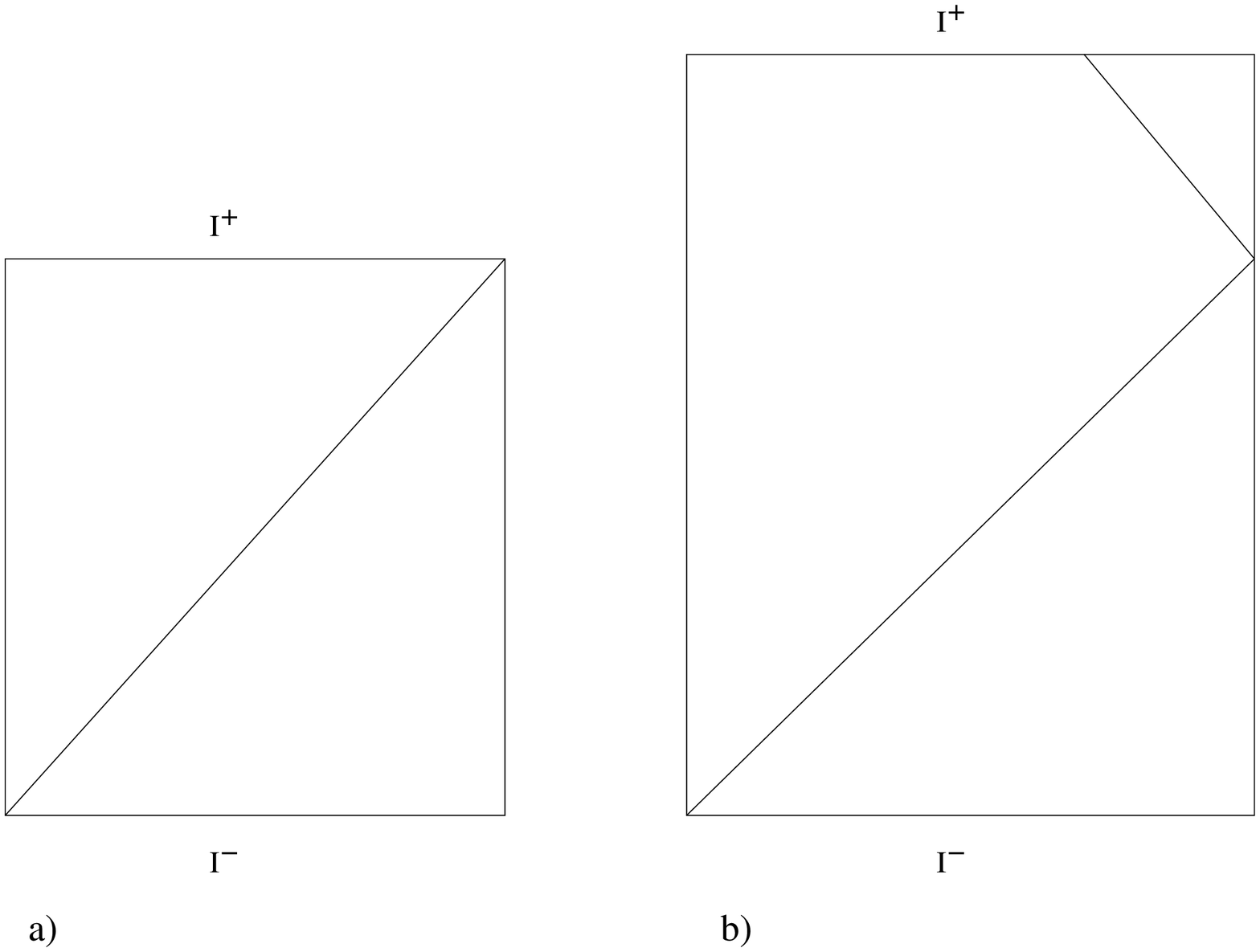,width=65mm}{Conformal diagrams associated
with a) pure dS space in global coordinates and b) the s7-- and
the s8--brane backgrounds. \label{penrose1}}

We assume that the s7--brane has a metric ansatz of the form
(\ref{metricsbrane}). The Friedmann constraint (\ref{thebig})
becomes \beq \rho = 28 \left[\left(\frac{\dot{a}}{a}\right)^{2} +
\frac{k}{a^{2}} \right], \eeq from which the scale factor $R(t)$
has dropped out. For simplicity we consider that the spatial
direction transverse to the unstable 8--brane is compactified on a
circle of fixed size.\footnote{The more general case with
time--dependent $R(t)$ will be treated in ref.~\cite{robbins}.}
The effective ($8+1$)--dimensional gravitational action is
(\ref{actionmain}) with $p=8$, $\psi=0$ and $a=-2$, supplemented
with the dimensionally reduced cosmological constant term
(\ref{cosmoterm}), \beq S_{\Lambda} = -\frac{1}{16\pi
G_{\scriptscriptstyle N}}\int d^{9}x \sqrt{-g} \Lambda_{b}. \eeq
The resulting system is simply gravity coupled to a scalar field
with a positive potential of the form \beq \label{newpot} V'(\phi)
= \Lambda_{b} + \frac{C^{2}}{2} e^{2\phi}. \eeq This corresponds
to the potential illustrated on figure~\ref{evolfig} with the sign
of $\alpha$ changed and the $\phi\rightarrow -\infty$ region
lifted to $+\Lambda_{b}$.

The gravitational backreaction of a scalar field rolling in a
potential of the form (\ref{newpot}) should lead to asymptotically
dS spacetimes. The $k=+1$ solutions correspond to a 9--dimensional
spacetime with a bounce separating asymptotic dS regions in the
past (including the conformal boundary $I^{-}$) and in the future
(with the boundary $I^{+}$). Figure~\ref{penrose1} shows both the
Penrose diagrams associated with a spherical s7--brane and with
pure dS space in global coordinates. We consider the $k=0$ at the
end of this section.

The Type~IIA s8--brane in the presence of D8--branes will lead to
spacetimes which are qualitatively similar to those associated
with the s7--branes. In Type~IIA we consider an unstable 9--brane
in the presence of $N$ D8--branes smeared along the spatial
transverse direction.

We now provide further support to the implicit assumption we made
that spacetimes supported by scalar fields rolling in potentials
of the form~(\ref{newpot}) lead to non--singular asymptotically dS
spacetimes. Concrete applications will be considered in
ref.~\cite{robbins}. In particular we study the $t>0$ region
following a bounce, the analysis being unchanged for $t<0$. For
the s7--brane ($p=8$) the acceleration of the scale factor is
provided by the expression \beq 8 \frac{\ddot{a}}{a} =
-\frac{1}{2}\dot{\phi}^{2} + \frac{1}{7}\left(
\frac{C^{2}}{2}e^{2\phi} + \Lambda_{b} \right). \eeq As pointed
out in section~\ref{sec3} the time--reversal symmetric initial
conditions at the bounce are $\dot{a}(0)=0=\dot{\phi}(0)$ and
$\phi(0)=\phi_{0}$. The size of the unstable brane when it bounces
is determined by solving the Friedmann constraint. There
will be a phase of positive acceleration for $t\geq 0$. The
question is whether of not it will last and, if not, whether the
spacetime will be driven to gravitational collapse.

Curves~II and IV on figure~\ref{violation} represent typical
effective equations of state associated with non--singular $k=+1$
solutions. A possibility is that the spacetimes of interest will have positive
acceleration at all times (curve~II). It is also conceivable that
the initial phase of positive acceleration is followed by a period
of negative acceleration where the kinetic energy of the scalar field (the only source
not violating the {\small SEC}) comes to dominate. In
section~\ref{sec3} we found many examples ($\Lambda_{b}=0$) for which the spacetime
never exits this phase (see curve~III). Then the dominating scalar
field contribution drives the spacetime into a phase contraction.
Since from that point on the sources violating the {\small SEC}
become increasingly negligible the singularity theorems apply and
predict that singular points will appear in the future. A point
we will be making is that for $\Lambda_{b}\neq 0$
contraction is unlikely to occur.

The equation of state for the scalar field associated with the
s7--brane is \beq \label{equationstate} w =
\frac{\dot{\phi}^{2}-\frac{C^{2}}{2}e^{2\phi}
-\Lambda_{b}}{\dot{\phi}^{2}+\frac{C^{2}}{2}e^{2\phi}
+\Lambda_{b}}.\eeq For $\Lambda_{b}=0$ the crux of the matter is
which component ($\dot{\phi}^{2}$ or $(C^{2}/2)e^{2\phi}$)
dominates for large $|\phi|$. If the kinetic energy term dominates
then the asymptotic equation of state is close to $w=1$ and, since
this violates the {\small SEC}, singular points appear in the
future. Non--singular solutions are those for which the potential
term dominates in such a way that the {\small SEC} is violated in
the future. This is clearly a very delicate system which appears
to always be on the verge of developing singularities.

For $\Lambda_{b}\neq 0$ the situation is different. Inspecting
eq.~(\ref{equationstate}) we see that as long as $\dot{\phi}$ is
kept small with respect to $\Lambda_{b}$ the kinetic term will
never dominate. It is possible that after the bounce ($t=0$,
$\dot{\phi}=0$) the spacetime will enter a phase where the {\small
SEC} is violated (negative acceleration) but it is likely to be
temporary and lead to a state with positive acceleration in the
asymptotic future. The gradient associated with the kinetic energy is \beq
\frac{d}{dt}\dot{\phi} = - \left( p\dot{\phi}\frac{\dot{a}}{a} +
C^{2} e^{2\phi}\right). \eeq For $\Lambda_{b}=0$ increasing $C$
favors the potential term in the equation of state
(\ref{equationstate}) but leads to an increase of $\ddot{\phi}$.
This is why the $\dot{\phi}^{2}$ tends to dominate anyway in the
future. However we see that for $\Lambda_{b}\neq 0$ the scalar
field does not couple to the cosmological constant.\footnote{This
is because the only potential--dependent expression appearing in
the dilaton equation of motion is $\partial V/\partial \phi$.}
Therefore by tuning the number of stable D--branes ($N$) it should
be possible to find solutions where the vaccum energy contribution
always dominates ($w=-1$) in the future.

For the $k=0$ solutions a bounce is not allowed by the Friedmann
constraint. The solutions are therefore either forever contracting
or forever expanding. The corresponding non--singular solutions
should be akin to the representation of dS space in terms of
expanding inflationary coordinates (the corresponding conformal
diagrams are described in ref.~\cite{myersds}). However we can
show that no non--singular solutions exist in this case which is
why they were not included in the conjecture. A very critical
constraint associated with the systems of interest is that they
respect the {\small WEC} ($w\geq -1$). For a scalar field
(relevant for our s7-- and s8--branes) this implies that \beq
\frac{d\rho}{dt} = - 2p \frac{\dot{a}}{a} \dot{\phi}^{2}. \eeq This
means the energy density, \beq \rho =
\frac{1}{2}\dot{\phi}^{2} + \frac{C^{2}}{2} e^{2\phi} +
\Lambda_{b}, \eeq must decrease (increase) during a phase of
expansion (contraction). However a non--singular $k=0$
asymptotically dS solution must be associated with \beq
\lim_{t\rightarrow \pm \infty} \left(\frac{1}{2}\dot{\phi}^{2} +
\frac{C^{2}}{2}e^{2\phi}\right) = 0. \eeq This last condition is
in direct contradiction with the {\small WEC}.

The {\small WEC} does not appear to put much constraint on the
$k=1$ solutions. However it will be important to verify that
consistent scalar field dynamics exist for the s7-- and s8--branes
solutions described in this section \cite{robbins}.

\subsubsection{The other s--branes}
\label{otherbranes}

We now briefly explain why we expect the conjecture
to hold for s--branes of all dimensions. We assume these
gravitational objects are associated with a metric ansatz of the
form (\ref{metricsbrane}) and that the volume of the
9--dimensional Cauchy surfaces goes through a bounce at $t=0$.
This region must be associated with a phase of positive
acceleration. This is possible as long as the boundary conditions
on the dilaton and the RR fields are chosen in such a way that the
cosmological constant term dominates (on the RHS of the $\ddot{a}/a$ expression)
around the bounce (clearly this
cannot happen when $\Lambda_{b}=0$). Another
condition at $t=0$ for a bounce to occur is
\beq p \frac{\dot{a}}{a} + (9-p) \frac{\dot{R}}{R} = 0. \eeq
Solutions with $\dot{a}(0)=0$ and $\dot{R}(0)=0$ can lead
to a bounce but it is not the only option. This is fortunate
because these boundary conditions exclude
s--branes with the symmetries $k_{1}=0$ and $k_{2}=-1$ proposed in
ref~{\cite{strominger1}} as is seen by inspection of the Friedmann
constraint (\ref{thebig}). In principle all combinations of
$k_{1}$ and $k_{2}$ are allowed given appropriate boundary
conditions for $\dot{a}/a$ and $\dot{R}/R$.

A lesson we have learned in this work is that generating a gravitational bounce
does not guarentee the whole spacetime is non--singular. In
fact we have shown in section~\ref{sec3} that when there is no vacuum energy more often
than not bouncing spacetimes are singular both in the past and the
future. Our claim here is that the presence of a
non--vanishing vacuum energy term (the stable branes) will resolve
these big--bang and big--crunch singularities. The dilaton and the
RR fields satisfy the {\small SEC}. We have shown in
section~\ref{sec1} that for large values of the scale factor (in
this case the average scale factor for the full ten--dimensional
spacetime) a cosmological constant term always comes to dominate
over contributions that respect the energy condition.

The dynamics of the hypothetical non--singular s--branes can be such that
the acceleration of the spacetime is always positive, \ie, the {\small SEC}
is violated for all times. Another possibility is that an
s--brane solution evolving from $t=0$ will traverse three phases.
First it will get out of its initial phase of positive
acceleration responsible for the bounce and enter a phase where
the dilaton and RR fields dominate. It is possible that the
spacetime will enter into a phase of contraction before the
cosmological constant term becomes dominant again. The singularity
theorems then predict singular points will develop in the future.
However there should be a range of initial conditions
allowing the s--branes to successfully enter a third phase which
is vacuum--dominated in the future.

The non--singular s--brane solutions discussed here would be
asymptotically dS$_{10}$.
The issues related to the non--singular nature of the
lower--dimensional s--branes will be fully addressed in
ref.~\cite{robbins}. We believe it is likely the conjecture
presented here will be proven. However a more conservative
statement at this stage is that the no--go theorems presented in
ref.~\cite{walcher,peet2} are completely evaded by s--branes when
they are cast in the more general and, perhaps, more realistic
context presented here.

\subsection{Open--closed duality and quantum effects}

Let us close this section by making a few comments regarding the
relevance of supergravity s--branes for the physics of unstable
D--branes. According to the open--closed string duality
conjectured by Sen \cite{openclosed} we should consider unstable
D--branes (at least at tree--level) from the point of view of either
closed strings or open strings, not both. Perhaps this can be interpreted to mean that
studying s--branes in a supergravity context with the massless
closed string modes sourced by the open string tachyon is not the
correct thing to do \cite{peet}. Another problem for the
supergravity approach is that as the open string tachyon rolls it
emits massive closed strings, the contribution of which is not
negligible \cite{liu}. For example unstable D0--branes decay
essentially entirely into massive closed strings. This result is
based on the tree--level calculation of the amplitude for the
emission of a closed string mode in a theory with appropriate
marginal tachyon boundary operator inserted. The integration over
kinetically allowed modes (\ie, those with $m^{2}\leq g_{s}^{-1}$)
leads to a divergent result. This is interpreted as tachyon matter
\cite{sen2} evaporation into massive closed strings. For unstable
$p$--branes with $p>1$ the presence of inhomogeneous tachyon
degrees of freedom is likely to affect significantly the physics
and results obtained assuming homogeneity must be used with
caution. It is nevertheless interesting to note that for $p>2$ the
integration over massive modes leads to finite results \cite{liu}.
However for branes with a characteristic size of the order
$a_{0}\sim l_{s}$ the total amplitude is again divergent while for
large enough branes the result is finite. We use this as a hint
that massive closed string modes emission can be made negligible
by considering unstable branes of appropriate size. This could, in
principle, justify using the supergravity approximation, \ie,
considering only the massless closed string modes when studying
these objects.

\section{Discussion}

Refs.~\cite{townsend1,townsend2,emparan} pointed out that positive
acceleration effects in ($3+1$)--dimensions can be obtained from
ten-- and eleven--dimensional supergravity. This led to a renewed
interest in cosmological solutions associated with the low~energy
dynamics of string theories
\cite{ulf,today,double,hyper,acceleration_positive}. Our analysis
is complementary and focuses on studying (and trying to resolve)
the pervading singularities associated with super--gravitational
time--dependent backgrounds. We examined the possibility of
obtaining bounces in theories associated with very simple
dimensional reductions of supergravity theories. Considering flux
compactifications on maximally symmetric Euclidean spaces we find
a negative result for bouncing solutions in the effective
($p+1$)--dimensional gravitational theory. For the non--singular
bounces presented in section~\ref{sec3} we found they cannot be
uplifted to ten or eleven dimensions. This does not exclude that
an embedding could be found in the context of a richer
compactification scheme, \ie, by considering transverse manifolds
associated with more interesting symmetries (see, \eg ,
ref.~\cite{ulf}). The only bouncing solutions we found that can be
embedded in higher--dimensional theories are those containing both
a big--crunch and a big--bang singularity. It is possible that one
of these singularities is only an illusion of the compactification
scheme used. In fact it is likely the dynamics of the breathing
mode will in effect resolve one of those singularities. The other
type of effective solutions considered is referred to as the
singular {\it infinite throat}. In particular let us consider the
big--crunch throat ($s=+1$ in section~\ref{sec2}) possessing a
curvature singularity in their future. For $p\leq 5$ ($d=10$) and
$p\leq 6$ ($d=11$), the singularity structure remains
qualitatively the same after the solution has been uplifted.
However for other values of $p$ the curvature singularity in the
future remains but the higher--dimensional spacetime becomes
singular in the past as well. This is an example of the phenomenon
by which the regular nature of the lower--dimensional spacetime is
only an illusion since singular points appear in the past as is
seen by performing a higher--dimensional analysis. It is therefore
clear that the concept of singularity resolution is ambiguous
when considering effective cosmological
solutions in the context of higher--dimensional theories.

So we provided an explicit example where a deceptively
non--singular region in ($p+1$) dimensions is in fact associated
with singular points if we consider its embedding in a
higher--dimensional theory. More generally it should be possible
to find effective theories (obtained from compactification of a
higher--dimensional theory) admitting non--singular
gravitational solutions containing a bounce. (The explicit
big--crunch example we provided can be regarded as exemplifying
the behavior in the past of such a solution.) Based on the
analysis performed in section~\ref{applications} we can
extrapolate what the higher--dimensional behavior of such a
solution might be. The cosmological singularity theorems predict
that the higher--dimensional spacetime must contain singular
points,\footnote{It is important to recall that we are assuming
the higher--dimensional theory (in this case supergravity)
satisfies the conditions in these theorems.} and we assume these
are in the past. For the future of these solutions there is really
nothing special to say. Whatever lower--dimensional cosmological
evolution is found should be what is detected by a
($p+1$)--dimensional observer.\footnote{There is of course the
issue of making sure that physical predictions are not affected
too much in the resulting theory if we consider $p=3$ spacetimes.
For example, the gravitational coupling must remain small enough
to conform with experiments \cite{ADD}. These issues were
consistently ignored throughout our work. Also, the comments in
this discussion are relevant in a braneworld'ish context where
matter as we know it is confined on the ($p+1$)--dimensional
manifold.} The past will be more subtle and interesting. The
($p+1$)--dimensional observer would observe in her past a bounce,
\ie, a region where its universe becomes very small. She would
probably conclude that this region is associated with singular
points where quantum gravitational effects become important.
However this is not the case assuming of course that the volume of
the higher--dimensional spacetime is large enough. This phenomenon
is an example where the dynamics of the transverse dimensions
resolve a spacelike singularity. Of course this state of affair
would only be temporary since the spacetime {\it must} contain
singular points further in the past based on the cosmological
singularity theorems. This situation might be interesting for
cosmology because it can push back in the past (perhaps very far)
the time where issues of quantum gravity become important. This,
for example, would be relevant if the breathing mode plays the
role of the inflaton. Of course up to now there was not
very much success in deriving potentials from string theory
leading to realistic models of inflation.

We have also considered a type of solutions corresponding to
Lorentzian wormholes. These are the static equivalent of bouncing
cosmologies. These geometries are associated with two disconnected
Lorentzian boundaries. An important ($3+1$)--dimensional result
found in refs.~\cite{thorne,visser} is that such spacetimes can
only exist in theories containing sources violating the {\small
WEC}. This result is obtained by performing a local analysis. When
global aspects are considered it is found that static spacetimes
with disconnected boundaries exist (even if the {\small WEC} is
satisfied) but that the boundaries must be separated by an
horizon. This conclusion is reached by proving topological
censorship theorems \cite{galloway}. A Schwarzschild black hole is
a spacetime with causally disconnected boundaries.

We generalized these results to higher--dimensional spacetimes and
found the main conclusion to be unchanged. In fact reproducing the
analysis found in section~\ref{bounces} for wormholes we find that
the {\small NEC} must be violated for traversable wormhole throats
(\ie, a spacetime with causally related disconnected boundaries)
to exist. However it is interesting to wonder whether or not it is
possible to find traversable wormhole geometries in the context of
($p+1$)--dimensional effective theories obtained by
compactification of higher--dimensional theories. The main
obstruction, as compared to the case of bouncing cosmologies, is
that it is typically much harder to violate the {\small NEC}
($\rho+p\geq 0$) than to violate the {\small SEC} ($(n-2)\rho + n
p \geq 0$). For example a scalar field $\phi$ in a potential
$V(\phi)$ is such that $\rho+p=\dot{\phi}^{2}/2$ which is clearly
positive--definite. A possible way to obtain lower--dimensional
scalar field models associated with traversable wormhole solutions
is if the resulting effective theory contains a curvature coupling
terms of the right sign. Even if we can find such effective
theories it is guaranteed, based on the topological theorems, that
an horizon will appear as seen from the higher--dimensional
spacetime point of view.

If they exist it would be very interesting to cast Lorentzian
wormholes in a gauge/gravity duality context. An example of
non--traversable wormhole is the AdS$_{3}$ eternal black hole. In
this case the conformal field theories on the two boundaries
(separated by an horizon) were argued in ref.~\cite{maldacena2} to
be in an entangled state. In an asymptotically AdS traversable
wormhole there would be two causally connected maximally symmetric
boundaries. The asymptotia would both have boundaries with
conformal group isometries. This is reminiscent of the
renormalization group flow interpretation of asymptotically
anti--de~Sitter spacetimes in AdS/CFT. However the situation would
be different since in the RG flow picture one of the fixed points
(IR) is not a boundary \cite{us}. The presence of two causally
connected boundaries should be closely related to cases where a
time--dependent background is conjectured to be dual to Euclidean
field theories on spacelike boundaries. An attempt to find such a
duality is the dS/CFT correspondence \cite{dscft}. In this case
the gravitational background is either pure or asymptotically dS
space \cite{dscft2,myersds}.

\acknowledgments I would like to thank
Christian~Armendariz--Picon, Sean~Carroll, Damien~Easson,
Aki~Hashimoto, Rajesh~Govindan, Stefan~Hollands, Alex~Maloney,
Shiraz~Minwalla, Vasilis~Niarchos, Kazumi~Okuyama, Amanda~Peet,
Daniel~Robbins, Sav~Sethi, Gary~Shiu and Robert~Wald for useful
conversations. Part of this research was supported by NSERC of
Canada.

\appendix

\small{
\section{Scalar field inter--breading}
\label{scalarstuff}

In section~\ref{applications} we considered $(p+1)$--dimensional
gravitational models obtained from compactification of
higher--dimensional theories. Only solutions with the dilaton
turned off were studied. Here we consider in some detail the case
where both the breathing mode and the dilaton are excited. The
effective potential is the of the form \beq V(\phi,\psi) =
\frac{C^{2}}{2} e^{-\left(a\phi +
\sqrt{\frac{2np^{2}}{(p-1)(n+p-1)}}\right)\psi} -\sigma n(n-1)
e^{-\sqrt{\frac{2(n+p-1)}{n(p-1)}}\psi}. \eeq The solutions with
$k=0$ (the spatial curvature on the ($p+1$)--dimensional
Lorentzian sub--manifold) and $\sigma=-1$ are the supergravity
s--brane solutions found in
refs.~\cite{strominger1,gutperle,myerssbrane}. Our interest lies
in finding ($p+1$)--dimensional solutions containing a bounce. We
have since in section~\ref{applications} that simple truncations
to one scalar field do not lead to regular solutions. In this
section we investigate whether or not the dilaton field can be
used to resolve the singularities.

We consider the simple case $\sigma=0$, \ie, the curvature of the
transverse dimensions vanishes. Using the metric ansatz
(\ref{hypermetric}) with the gauge $A=pB$ the equations of motion
are \beq \label{eom11} \ddot{B} = \frac{1}{p-1}e^{2pB - \alpha\psi
-a\phi} - k(p-1)e^{2(p-1)B}, \eeq \beq \label{eom22} \ddot{\phi} =
a e^{2pB-\alpha\psi-a\phi}, \eeq \beq \label{eom33} \ddot{\psi} =
\alpha e^{2pB-\alpha\psi-a\phi}, \eeq where $\alpha =
\sqrt{\frac{2np^{2}}{(p-1)(n+p-1)}}$ and $a=(4-p)/2$. The
Friedmann constraint takes the form \beq \frac{p(p-1)}{2}\left(
\dot{B}^{2} + ke^{2(p-1)B}\right) = \frac{1}{4}\left(
\dot{\phi}^{2} + \dot{\psi}^{2} \right) + \frac{C^{2}}{4}
e^{2pB-\alpha\psi-a\phi}. \eeq

By inspecting eqs.~(\ref{eom22}) and (\ref{eom33}) we can write
down the solution for the dilaton in terms of the breathing mode,
\beq \phi(t) = \frac{a}{\alpha} \psi(t) + c_{2}t + c_{1}, \eeq
where $c_{1}$ and $c_{2}$ are constants of integration. Using this
information the system of differential equations we need to solve
takes the simpler form \beq \ddot{B} =
\frac{\bar{C}^{2}}{2(p-1)}e^{2pB - \bar{\alpha}\psi - a c_{1}t} -
k(p-1) e^{2(p-1)B},  \eeq \beq \ddot{\psi} =
\frac{\alpha\bar{C}^{2}}{2} e^{2pB - \bar{\alpha}\psi - a c_{1}t},
\eeq and the Friedmann constraint becomes \beq
\frac{p(p-1)}{2}\left( \dot{B}^{2} + k e^{2(p-1)B}\right) =
\frac{1}{4}\left( \dot{\phi}^{2} +\dot{\psi}^{2}\right) +
\frac{\bar{C}^{2}}{4} e^{2pB - \bar{\alpha}\psi - a c_{1}t},  \eeq
where we have defined \beq \bar{\alpha} =
\frac{a^{2}+\alpha^{2}}{\alpha}, \eeq and \beq \bar{C}^{2} = C^{2}
e^{-ac_{2}}. \eeq Similarly to what we did in
section~\ref{applications} we consider bouncing spacetimes with
the boundary conditions \beq \dot{\psi}(0) = 0 = \dot{B}(0),
\;\;\; \psi(0) = \psi_{0}. \eeq This implies that the boundary
conditions for the dilaton are of the form \beq c_{2} = \phi(0) -
\frac{a}{\alpha} \psi_{0}, \;\;\; c_{1}=\dot{\phi}(0). \eeq There
are two cases potentially leading to bouncing spacetimes. The
simplest one consists in considering that the dilaton bounces
simultaneously with the breathing mode and the gravitational
field. This corresponds to the system with $c_{1}=0$. However we
have already treated this case in section~\ref{sec3}. We can
therefore use the results found there by simply replacing $\alpha$
with $\bar{\alpha}$ and $C$ with $\bar{C}$. We found earlier that
all values of $\alpha$ obtained from string compactifications are
such that $\alpha<\alpha_{c}$ and therefore lead to singular
cosmologies. Since we always have $\bar{\alpha}>\alpha$ the same
results apply to this simple dilaton--breathing mode system.

A potentially more interesting case consists in allowing for the
kinetic energy stored in the dilaton field to be non--vanishing at
the hypothetical bounce, \ie, $c_{1}\neq 0$. The Friedmann
constraint at $t=0$ then takes the form \beq c_{1}^{2} = 2
e^{2pB(0)} \left( p(p-1) - \frac{\bar{C}^{2}}{2} e^{-\bar{\alpha
\psi(0)}} \right). \eeq We have performed numerous numerical
experiments and our non--definite prediction is that the dilaton
does not resolve the singularities associated with the bouncing
spacetimes.

}
{}



\begin{thebibliography}{}

\bibitem{supernovae} S.~Perlmutter {\it et al.}  [Supernova Cosmology Project Collaboration],
``Measurements of the Cosmological Parameters $\Omega$ and
$\Lambda$ from the First Seven Supernovae at z$\ge$0.35,''
Astrophys.\ J.\  {\bf 483}, 565 (1997) [arXiv:astro-ph/9608192];\\
S.~Perlmutter {\it et al.}  [Supernova Cosmology Project
Collaboration], ``Discovery of a Supernova Explosion at Half the
Age of the Universe and its Cosmological Implications,''
Nature {\bf 391}, 51 (1998) [arXiv:astro-ph/9712212];\\
A.G.~Riess {\it et al.}  [Supernova Search Team Collaboration],
``Observational Evidence from Supernovae for an Accelerating
Universe and a Cosmological Constant,''
Astron.\ J.\  {\bf 116}, 1009 (1998) [arXiv:astro-ph/9805201].;\\
N.A.~Bahcall, J.P.~Ostriker, S.~Perlmutter and P.J.~Steinhardt,
``The Cosmic Triangle: Revealing the State of the Universe,''
Science {\bf 284}, 1481 (1999) [arXiv:astro-ph/9906463].

\bibitem{kklt}
S.~Kachru, R.~Kallosh, A.~Linde and S.~P.~Trivedi, ``De Sitter
vacua in string theory,'' Phys.\ Rev.\ D {\bf 68}, 046005 (2003)
[arXiv:hep-th/0301240].

\bibitem{moore}
H.~Liu, G.~Moore and N.~Seiberg, ``Strings in a time-dependent
orbifold,'' JHEP {\bf 0206}, 045 (2002) [arXiv:hep-th/0204168];
H.~Liu, G.~Moore and N.~Seiberg, ``Strings in time-dependent
orbifolds,'' JHEP {\bf 0210}, 031 (2002) [arXiv:hep-th/0206182].

\bibitem{chicago}
B.~Craps, D.~Kutasov and G.~Rajesh, ``String propagation in the
presence of cosmological singularities,'' JHEP {\bf 0206}, 053
(2002) [arXiv:hep-th/0205101];
M.~Berkooz, B.~Craps, D.~Kutasov and G.~Rajesh, ``Comments on
cosmological singularities in string theory,'' JHEP {\bf 0303},
031 (2003) [arXiv:hep-th/0212215].

\bibitem{coleman}
S.~R.~Coleman, ``Why There Is Nothing Rather Than Something: A
Theory Of The Cosmological Constant,'' Nucl.\ Phys.\ B {\bf 310},
643 (1988).

\bibitem{klebanov}
I.~R.~Klebanov, L.~Susskind and T.~Banks, ``Wormholes And The
Cosmological Constant,'' Nucl.\ Phys.\ B {\bf 317}, 665 (1989).

\bibitem{giddings}
S.~B.~Giddings and A.~Strominger, ``String Wormholes,'' Phys.\
Lett.\ B {\bf 230}, 46 (1989).

\bibitem{maoz}
J.~Maldacena and L.~Maoz, ``Wormholes in AdS,''
arXiv:hep-th/0401024.

\bibitem{mcinnes}
B.~McInnes, ``Quintessential Maldacena-Maoz cosmologies,''
arXiv:hep-th/0403104.

\bibitem{thorne}
M.~S.~Morris and K.~S.~Thorne, ``Wormholes In Space-Time And Their
Use For Interstellar Travel: A Tool For Teaching General
Relativity,'' Am.\ J.\ Phys.\  {\bf 56}, 395 (1988).

\bibitem{visser}
M.~Visser, S.~Kar and N.~Dadhich, ``Traversable wormholes with
arbitrarily small energy condition violations,'' Phys.\ Rev.\
Lett.\  {\bf 90}, 201102 (2003) [arXiv:gr-qc/0301003];
C.~Barcelo and M.~Visser, ``Scalar fields, energy conditions, and
traversable wormholes,'' Class.\ Quant.\ Grav.\  {\bf 17}, 3843
(2000) [arXiv:gr-qc/0003025];
D.~Hochberg and M.~Visser, ``The null energy condition in dynamic
wormholes,'' Phys.\ Rev.\ Lett.\  {\bf 81}, 746 (1998)
[arXiv:gr-qc/9802048];
D.~Hochberg and M.~Visser, ``Dynamic wormholes, anti-trapped
surfaces, and energy conditions,'' Phys.\ Rev.\ D {\bf 58}, 044021
(1998) [arXiv:gr-qc/9802046];
M.~Visser and D.~Hochberg, ``Generic wormhole throats,''
arXiv:gr-qc/9710001;
M.~Visser, ``Traversable wormholes: The Roman ring,'' Phys.\ Rev.\
D {\bf 55}, 5212 (1997) [arXiv:gr-qc/9702043];
M.~Visser, ``Traversable Wormholes: Some Simple Examples,'' Phys.\
Rev.\ D {\bf 39}, 3182 (1989).

\bibitem{galloway}
G.~J.~Galloway, K.~Schleich, D.~M.~Witt and E.~Woolgar,
``Topological censorship and higher genus black holes,'' Phys.\
Rev.\ D {\bf 60}, 104039 (1999) [arXiv:gr-qc/9902061];
``The AdS/CFT correspondence conjecture and topological
censorship,'' Phys.\ Lett.\ B {\bf 505}, 255 (2001)
[arXiv:hep-th/9912119].

\bibitem{strominger1}
M.~Gutperle and A.~Strominger, ``Spacelike branes,'' JHEP {\bf
0204}, 018 (2002) [arXiv:hep-th/0202210].

\bibitem{gutperle}
C.~M.~Chen, D.~V.~Gal'tsov and M.~Gutperle, ``S-brane solutions in
supergravity theories,'' Phys.\ Rev.\ D {\bf 66}, 024043 (2002)
[arXiv:hep-th/0204071].

\bibitem{myerssbrane}
M.~Kruczenski, R.~C.~Myers and A.~W.~Peet, ``Supergravity
S-branes,'' JHEP {\bf 0205}, 039 (2002) [arXiv:hep-th/0204144].

\bibitem{buchel1}
A.~Buchel, P.~Langfelder and J.~Walcher, ``Does the tachyon
matter?,'' Annals Phys.\  {\bf 302}, 78 (2002)
[arXiv:hep-th/0207235].

\bibitem{peet}
F.~Leblond and A.~W.~Peet, ``SD-brane gravity fields and rolling
tachyons,'' JHEP {\bf 0304}, 048 (2003) [arXiv:hep-th/0303035].

\bibitem{alex}
G.~Jones, A.~Maloney and A.~Strominger, ``Non-Singular Solutions
for S-branes,'' arXiv:hep-th/0403050.

\bibitem{pbranes}
G.~T.~Horowitz and A.~Strominger, ``Black Strings And P-Branes,''
Nucl.\ Phys.\ B {\bf 360}, 197 (1991); C.~M.~Hull and
P.~K.~Townsend, ``Unity of superstring dualities,'' Nucl.\ Phys.\
B {\bf 438}, 109 (1995) [arXiv:hep-th/9410167]; P.~K.~Townsend,
``The eleven-dimensional supermembrane revisited,'' Phys.\ Lett.\
B {\bf 350}, 184 (1995) [arXiv:hep-th/9501068]; E.~Witten,
Nucl.\ Phys.\ B {\bf 443}, 85 (1995) [arXiv:hep-th/9503124];
\bibitem{Strominger:1995cz}
A.~Strominger, ``Massless black holes and conifolds in string
theory,'' Nucl.\ Phys.\ B {\bf 451}, 96 (1995)
[arXiv:hep-th/9504090].

\bibitem{mtw}
C.~W.~Misner, K.~S.~Thorne and J.~A.~Wheeler, ``Gravitation'',
W.~H.~Freeman and Company, 1973.

\bibitem{wald}
R.~M.~Wald, ``General Relativity,'' The University of Chicago
Press, 1984.

\bibitem{hawking}
S.~W.~Hawking and G.~F.~R.~Ellis, ``The large scale structure of
spacetime'', Cambridge University Press, 1973.

\bibitem{volovich}
M.~Spradlin, A.~Strominger and A.~Volovich, ``Les Houches lectures
on de Sitter space,'' arXiv:hep-th/0110007.

\bibitem{sen1}
A.~Sen, ``Non-BPS states and branes in string theory,''
arXiv:hep-th/9904207.

\bibitem{sen2}
A.~Sen, ``Rolling tachyon,'' JHEP {\bf 0204}, 048 (2002)
[arXiv:hep-th/0203211].
A.~Sen, ``Tachyon matter,'' JHEP {\bf 0207}, 065 (2002)
[arXiv:hep-th/0203265].

\bibitem{kutasov}
D.~Kutasov and V.~Niarchos, ``Tachyon effective actions in open
string theory,'' Nucl.\ Phys.\ B {\bf 666}, 56 (2003)
[arXiv:hep-th/0304045].

\bibitem{burgess}
C.~P.~Burgess, F.~Quevedo, R.~Rabadan, G.~Tasinato and I.~Zavala,
``Bouncing branes,'' arXiv:hep-th/0310122.

\bibitem{townsend1}
P.~K.~Townsend, ``Cosmic acceleration and M-theory,''
arXiv:hep-th/0308149.

\bibitem{townsend2}
P.~K.~Townsend and M.~N.~R.~Wohlfarth, ``Accelerating cosmologies
from compactification,'' Phys.\ Rev.\ Lett.\  {\bf 91}, 061302
(2003) [arXiv:hep-th/0303097].

\bibitem{emparan}
R.~Emparan and J.~Garriga, ``A note on accelerating cosmologies
from compactifications and S-branes,'' JHEP {\bf 0305}, 028 (2003)
[arXiv:hep-th/0304124].

\bibitem{ulf}
E.~Bergshoeff, A.~Collinucci, U.~Gran, M.~Nielsen and D.~Roest,
``Transient quintessence from group manifold reductions or how all
roads lead to Rome,'' arXiv:hep-th/0312102.

\bibitem{today}
L.~Jarv, T.~Mohaupt and F.~Saueressig, ``Quintessence Cosmologies
with a Double Exponential Potential,'' arXiv:hep-th/0403063.

\bibitem{sen10}
A.~Sen, ``Remarks on tachyon driven cosmology,''
arXiv:hep-th/0312153.

\bibitem{easson}
R.~Brandenberger, D.~A.~Easson and D.~Kimberly, ``Loitering phase
in brane gas cosmology,'' Nucl.\ Phys.\ B {\bf 623}, 421 (2002)
[arXiv:hep-th/0109165].

\bibitem{carroll}
S.~M.~Carroll, J.~Geddes, M.~B.~Hoffman and R.~M.~Wald,
``Classical stabilization of homogeneous extra dimensions,''
Phys.\ Rev.\ D {\bf 66}, 024036 (2002) [arXiv:hep-th/0110149].

\bibitem{walcher}
A.~Buchel and J.~Walcher, ``Comments on supergravity description
of S-branes,'' JHEP {\bf 0305}, 069 (2003) [arXiv:hep-th/0305055].

\bibitem{peet2}
F.~Leblond and A.~W.~Peet, ``A note on the singularity theorem for
supergravity SD-branes,'' arXiv:hep-th/0305059.

\bibitem{wang}
J.~E.~Wang, ``Twisting S-branes,'' arXiv:hep-th/0403094.

\bibitem{cliffb}
G.~Tasinato, I.~Zavala, C.P.~Burgess, F.~Quevedo, ``Regular
S-Brane Backgrounds,'' arXiv:hep-th/0403156.

\bibitem{robbins}
F.~Leblond and D.~Robbins, work in progress.

\bibitem{kazumi}
K.~Okuyama, ``Wess-Zumino term in tachyon effective action,'' JHEP
{\bf 0305}, 005 (2003) [arXiv:hep-th/0304108].

\bibitem{polchinski}
J.~Polchinski, ``String Theory. Vol. 2: Superstring Theory And
Beyond,'' Cambridge University Press, 1998.

\bibitem{kofman}
G.~N.~Felder, L.~Kofman and A.~Starobinsky, ``Caustics in tachyon
matter and other Born-Infeld scalars,'' JHEP {\bf 0209}, 026
(2002) [arXiv:hep-th/0208019];
G.~N.~Felder and L.~Kofman, ``Inhomogeneous fragmentation of the
rolling tachyon,'' arXiv:hep-th/0403073.

\bibitem{openclosed}
A.~Sen, ``Open-closed duality at tree level,'' Phys.\ Rev.\ Lett.\
{\bf 91}, 181601 (2003) [arXiv:hep-th/0306137]; ``Open-closed
duality: Lessons from matrix model,'' arXiv:hep-th/0308068.

\bibitem{liu}
N.~Lambert, H.~Liu and J.~Maldacena, ``Closed strings from
decaying D-branes,'' arXiv:hep-th/0303139.

\bibitem{double}
L.~Jarv, T.~Mohaupt and F.~Saueressig, ``Quintessence cosmologies
with a double exponential potential,'' arXiv:hep-th/0403063.

\bibitem{hyper}
I.~P.~Neupane, ``Accelerating cosmologies from exponential
potentials,'' arXiv:hep-th/0311071;
I.~P.~Neupane, ``Cosmic acceleration and M theory cosmology,''
arXiv:hep-th/0402021.

\bibitem{acceleration_positive}
 M.~N.~R.~Wohlfarth, ``Accelerating cosmologies and a phase transition in M-theory,''
Phys.\ Lett.\ B {\bf 563}, 1 (2003) [arXiv:hep-th/0304089];
N.~Ohta, ``A study of accelerating cosmologies from superstring/
M~theories,'' Prog.\ Theor.\ Phys.\  {\bf 110}, 269 (2003)
[arXiv:hep-th/0304172];
C.~M.~Chen, P.~M.~Ho, I.~P.~Neupane and J.~E.~Wang, ``A note on
acceleration from product space compactification,'' JHEP {\bf
0307}, 017 (2003) [arXiv:hep-th/0304177]; B.~McInnes, ``The strong
energy condition and the S-brane singularity problem,'' JHEP {\bf
0306}, 043 (2003) [arXiv:hep-th/0305107];
M.~Ito, ``On the solutions to accelerating cosmologies,'' JCAP
{\bf 0309}, 003 (2003) [arXiv:hep-th/0305130];
C.~M.~Chen, P.~M.~Ho, I.~P.~Neupane, N.~Ohta and J.~E.~Wang,
``Hyperbolic space cosmologies,'' JHEP {\bf 0310}, 058 (2003)
[arXiv:hep-th/0306291];
I.~P.~Neupane, ``Inflation from string/M-theory
compactification?,'' Nucl.\ Phys.\ Proc.\ Suppl.\  {\bf 129}, 800
(2004) [arXiv:hep-th/0309139];
M.~N.~R.~Wohlfarth, ``Inflationary cosmologies from
compactification,'' Phys.\ Rev.\ D {\bf 69}, 066002 (2004)
[arXiv:hep-th/0307179];
V.~D.~Ivashchuk, V.~N.~Melnikov and A.~B.~Selivanov,
``Cosmological solutions in multidimensional model with multiple
exponential potential,'' JHEP {\bf 0309}, 059 (2003)
[arXiv:hep-th/0308113].

\bibitem{ADD}
N.~Arkani-Hamed, S.~Dimopoulos and G.~R.~Dvali, ``The hierarchy
problem and new dimensions at a millimeter,'' Phys.\ Lett.\ B {\bf
429}, 263 (1998) [arXiv:hep-ph/9803315].
I.~Antoniadis, N.~Arkani-Hamed, S.~Dimopoulos and G.~R.~Dvali,
``New dimensions at a millimeter to a Fermi and superstrings at a
TeV,'' Phys.\ Lett.\ B {\bf 436}, 257 (1998)
[arXiv:hep-ph/9804398].

\bibitem{maldacena2}
J.~M.~Maldacena, ``Eternal black holes in Anti-de-Sitter,'' JHEP
{\bf 0304}, 021 (2003) [arXiv:hep-th/0106112].

\bibitem{us}
H.~Firouzjahi and F.~Leblond, ``The clash between de Sitter and
anti-de Sitter space,'' JCAP {\bf 0306}, 003 (2003)
[arXiv:hep-th/0209248].

\bibitem{dscft}
A.~Strominger, ``The dS/CFT correspondence,'' JHEP {\bf 0110}, 034
(2001) [arXiv:hep-th/0106113].

\bibitem{dscft2}
A.~Strominger, ``Inflation and the dS/CFT correspondence,'' JHEP
{\bf 0111}, 049 (2001) [arXiv:hep-th/0110087].

\bibitem{myersds}
F.~Leblond, D.~Marolf and R.~C.~Myers, ``Tall tales from de Sitter
space. I: Renormalization group flows,'' JHEP {\bf 0206}, 052
(2002) [arXiv:hep-th/0202094].


\end{thebibliography}
\end{document}